\documentclass[twocolumn]{article}
\usepackage[T1]{fontenc}

\usepackage[a4paper, margin=1in]{geometry}
\usepackage{microtype}
\usepackage{graphicx}
\usepackage{subfigure}
\usepackage{booktabs}
\usepackage{multirow}
\usepackage{authblk}
\usepackage{hyperref}
\usepackage[table]{xcolor}
\usepackage{xspace}
\usepackage{amsmath, amssymb, mathtools, amsthm}
\usepackage[capitalize,noabbrev]{cleveref}
\usepackage{pifont}

\usepackage{amsmath,amsthm,amsfonts,bm}

\newtheorem{definition}{Definition}

\def\eqref#1{equation~\ref{#1}}
\def\1{\bm{1}}

\def\vf{{\bm{f}}}

\def\vh{{\bm{h}}}

\def\vk{{\bm{k}}}

\def\vv{{\bm{v}}}

\def\vx{{\bm{x}}}

\DeclareMathAlphabet{\mathsfit}{\encodingdefault}{\sfdefault}{m}{sl}
\SetMathAlphabet{\mathsfit}{bold}{\encodingdefault}{\sfdefault}{bx}{n}

\def\gC{{\mathcal{C}}}
\def\gD{{\mathcal{D}}}
\def\gE{{\mathcal{E}}}

\def\gL{{\mathcal{L}}}

\def\gR{{\mathcal{R}}}

\def\gT{{\mathcal{T}}}

\def\gV{{\mathcal{V}}}

\def\sN{{\mathbb{N}}}

\def\sR{{\mathbb{R}}}

\def\sT{{\mathbb{T}}}

\usepackage{hyperref}
\hypersetup{
    colorlinks,
    linkcolor={red!50!black},
    citecolor={blue!50!black},
    urlcolor={blue!80!black}
}
\usepackage{natbib} 
\definecolor{mycustomcolor}{HTML}{e8f8f5}
\definecolor{mycustomcolor2}{HTML}{fef5e7}

\newcommand{\changed}[1]{\textcolor{black}{#1}}
\newcommand{\valstd}[2]{#1 {\scriptsize $\pm$ #2}}
\newcommand{\valstdf}[2]{\textbf{#1} {\scriptsize $\pm$ #2}}

\newcommand{\val}[1]{#1}
\newcommand{\valf}[1]{\textbf{#1}}

\newcommand{\xgb}{\textsc{LightGBM}\xspace}
\newcommand{\rdl}{\textsc{Rdl}\xspace}
\newcommand{\rgcn}{\textsc{HGSage}\xspace}
\newcommand{\our}{\textsc{LightRdl}\xspace}

\theoremstyle{plain}

\theoremstyle{definition}

\theoremstyle{remark}

\usepackage{titlesec}

\titlespacing*{\section}{0pt}{1.5ex plus .5ex minus .2ex}{0.8ex plus .2ex}
\titlespacing*{\subsection}{0pt}{1.2ex plus .5ex minus .2ex}{0.6ex plus .2ex}

\title{Boosting Relational Deep Learning with Pretrained Tabular Models}

\author[1]{Veronica Lachi$^*$}
\author[2]{Antonio Longa\thanks{Equal contribution}}
\author[3]{Beatrice Bevilacqua}
\author[1]{Bruno Lepri}
\author[2]{Andrea Passerini}
\author[3]{Bruno Ribeiro}

\affil[1]{Fondazione Bruno Kessler, Trento, Italy}
\affil[2]{Trento University, Trento, Italy}
\affil[3]{Purdue University, West Lafayette, USA}

\date{}

\begin{document}

\maketitle

\begin{abstract}
Relational databases, organized into tables connected by primary-foreign key relationships, are a common format for organizing data. Making predictions on relational data often involves transforming them into a flat tabular format through table joins and feature engineering,  which serve as input to tabular methods.  However, designing features that fully capture complex relational patterns remains challenging. Graph Neural Networks (GNNs) offer a compelling alternative by inherently modeling these relationships, but their time overhead during inference limits their applicability for real-time scenarios.
In this work, we aim to bridge this gap by leveraging existing feature engineering efforts to enhance the efficiency of GNNs in relational databases. Specifically, we use GNNs to capture complex relationships within relational databases—patterns that are difficult to featurize, while employing engineered features to encode temporal information, thereby avoiding the need to retain the entire historical graph and enabling the use of smaller, more efficient graphs.
Our \our approach not only improves efficiency, but also outperforms existing models. Experimental results on the RelBench benchmark demonstrate that our framework achieves up to $33\%$ performance improvement and a $526\times$ inference speedup compared to GNNs, making it highly suitable for real-time inference.

\end{abstract}

\section{Introduction}

Relational databases are extensively used in industry due to their flexibility, extensibility, and speed~\citep{berg2013history,johnson2016mimic,halpin2010information}.
The information is organized into tables and records entities, their features, relations (via primary and foreign key relationships), and events, such as transactions with their associated timestamps.
This format simplifies data maintenance and optimization while improving accessibility and retrieval through query languages like SQL \citep{codd1970relational,chamberlin1974sequel}. 
As a result, relational databases are integral to several applications, ranging from e-commerce platforms~\citep{agrawal2001storage} and social media networks~\citep{almabdy2018comparative} to banking systems \citep{aditya2002banks} and healthcare services \citep{park2014graph}. 
Due to the presence of both timestepped events and relations, tasks over relational databases tend to be both temporal and relational, such as forecasting future product sales and predicting future customer purchases and churn~\citep{robinson2024relbench}. 

For decades, companies have built in-house predictive models over these databases by flattening the complex temporal-relational data into tabular formats via meticulously engineered temporal and relational features~\citep{dong2018feature,ganguli2020machine}.
These flattened data, often referred to as {\em tabular data}, are then fed as input to tabular models such as \textsc{XGBoost} and \xgb \citep{ke2017lightgbm,chen2016xgboost}. 
However, capturing all complex relational patterns in the data through feature engineering can be challenging~\citep{lam2017one,zheng2018feature}.
A different paradigm proposes {\em neural network models}, such as Relational Graph Neural Networks (R-GNNs)~\citep{cvitkovic2020supervised,fey2024position,villaizan2024graph},  which model relational data directly as heterogeneous graphs. In this paradigm, each table row corresponds to an attributed node, and edges are defined through primary-foreign key relationships. R-GNNs eliminate the need for feature engineering~\citep{fey2024position} and have demonstrated success in simplifying predictive models~\citep{robinson2024relbench}. 

\begin{figure*}[t]
    \centering
    \includegraphics[width=\linewidth]{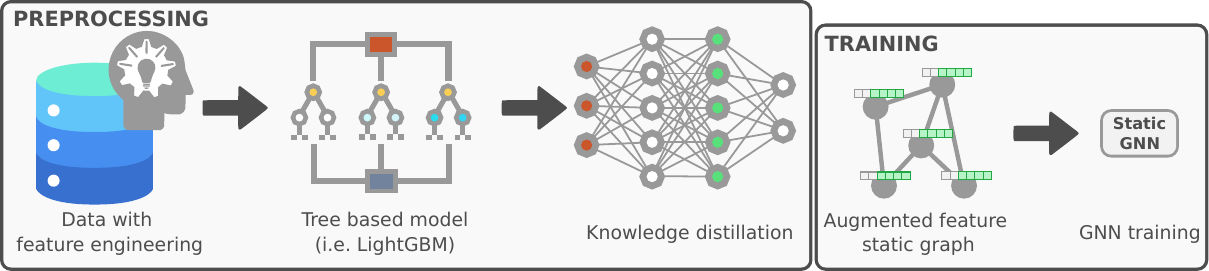}
    \caption{Overview of our proposed hybrid modeling framework \our. The pipeline begins with feature-engineered tabular data processed by a tree-based model (e.g., \xgb). Knowledge distillation then generates embeddings summarizing temporal information,  which are then utilized as additional node features for the static GNN responsible for the predictions.}
    \label{fig:diagram}
\end{figure*}

Despite the promise of R-GNNs, the strong performance and fast inference times of legacy tabular models \changed{— which we refer to as {\em pretrained tabular models}, that is, models trained on features engineered by domain experts —} remains a challenge for the widespread adoption of R-GNNs. These pretrained models are lightweight, benefiting from powerful features derived through extensive feature engineering, and they achieve fast inference speed that R-GNNs often struggle to match. Additionally, while R-GNNs can offer improved performance in certain scenarios, the gains can be modest, and in most cases, their results align with those of traditional tabular models (as evidenced in \citet{robinson2024relbench} and \Cref{tab:comparison}). This diminishes the incentive for transitioning to R-GNNs, particularly when inference speed is a priority.

This raises a key question: 
{\em How can we combine the efficiency of pretrained tabular models with the relational modeling power of R-GNNs to achieve both high accuracy and fast inference? Specifically,
can we leverage the feature engineering from pretrained tabular models to speed up R-GNNs while improving their performance? }

In this work, we propose a hybrid tabular-R-GNN modeling framework for predictive tasks on relational databases, which combines the relational modeling power of R-GNNs with the inference speed of tabular methods.
Our approach uses the knowledge extracted from pretrained tabular models, distilled into MLPs. These MLP-generated embeddings are integrated as additional features into an R-GNN. Incorporating these features allows us to train the R-GNN on a smaller graph constructed only from the immediate time period preceding the inference step, rather than on the entire historical graph required by R-GNNs. This enables us to effectively leverage the temporal dynamics handled by the tabular models, significantly reducing inference times. An overview of our method, which we call \our, is illustrated in \Cref{fig:diagram}.

Our \our not only enhances efficiency, but also outperforms existing models. Experimental results on the RelBench benchmark \citep{robinson2024relbench} demonstrate that our framework achieves up to $33\%$ performance improvement and $526\times$ inference speedup compared to R-GNNs, making it highly suitable for real-time inference.

\section{Learning on Relational Databases}

Relational databases organize data across multiple interconnected tables, each containing entities with a shared schema. Relationships between tables define dependencies among entities. Formally:

\begin{definition}[Relational Database]\label{def:rdb}
A {\em relational database} $\gR$ with $N$ tables consists of a set of entities $v \in \gV$, where each entity is uniquely indexed. The $i$-th table $\sT_i$ is defined as $\sT_i =\{ \vv_v \mid v \in \gV : \phi(v) = i\}$, where $\phi: \mathcal{V} \rightarrow [N]$ maps entities to their corresponding tables, and each row $\vv_v = (p_v, \vk_v, \vx_v, t_v)$ contains:
\begin{itemize}
    \item $p_v \in \sN$: the \textbf{primary key}, uniquely identifying $v$;
    \item $\vk_v = (k_v^1, \ldots, k_v^N)$: \textbf{foreign keys}, $k_v^c \in \sN$, linking to other tables, with $k_v^c = 0$ if no link exists to table $c$;
    \item $\vx_v = (x_v^1, \ldots, x_v^{d_i})$: \textbf{features};
    \item $t_v$: the \textbf{timestamp} indicating when $v$ appears.
\end{itemize}
The set of all tables is $\gT = \{ \sT_1, \ldots, \sT_N \}$. A link $(i, j) \in \gL$ exists between tables $i$ and $j$ if a foreign key in $\sT_i$ matches a primary key in $\sT_j$, i.e., $\gL = \{(i,j) \mid \exists u \in \sT_i, v \in \sT_j, k^j_u = p_v\}$.
\end{definition}

For instance, the \texttt{rel-hm} database from the RelBench benchmark~\citep{robinson2024relbench} captures customer purchase histories on the H\&M e-commerce platform. Further details about this dataset are provided in Appendix \ref{app:dataset}. The dataset includes:
(i) a \textbf{customer} table with attributes like gender and birth year;
(ii) a \textbf{product} table with product details such as description and size;
(iii) a \textbf{transaction} table recording customer purchases. 

Many real-world machine learning tasks on relational databases consist on predicting the future state of specific entities. 
For instance, on \texttt{rel-hm} one of the key tasks is forecasting the total sales of an article for the upcoming week.
This is relevant to H\&M as it enables effective inventory management, optimizes stock replenishment, and helps in crafting targeted marketing strategies to maximize revenue and reduce the risk of article shortages or overstock situations.
While some predictive tasks also focus on relationship predictions, such as forecasting interactions between entities, in this work we exclusively address entity-level tasks. Predictive tasks require the specification of a {\em seed time}, which is defined as ``the present'' in the prediction task. Consider the task of forecasting total sales of a product $v$. Given a {\em seed time} $t$ in days (``the present''), we would like to predict the sales of $v$ over the ``next week'', i.e., in the interval $[t, t+7]$ in the database. The {\em seed time} then also defines the training data {\em we can use}, which includes all database records with timestamps $t' < t$.

Addressing such tasks requires machine learning models to effectively handle the heterogeneity of the features, the non-linear relationships and the time, ensuring scalability during both training and inference.
The two most used approaches for tackling these challenges are tabular methods and graph-based models, which we discuss in the following.

\subsection{Tabular Methods}\label{subsec:treebasedmodels}
The most commonly used tabular methods for relational databases are gradient-boosted decision trees (GBDT)~\citep{gorishniy2021revisiting,shwartz2022tabular}. When applied to relational databases, these models require transforming relational data into a flat tabular format. This transformation often involves techniques such as joining tables or more sophisticated approaches such as feature engineering, where a domain expert creates a single table by manually designing features that encode relationships and aggregations from the relational database~\citep{heaton2016empirical}. Over the years, companies have heavily invested in developing ideal feature sets tailored to these methods, making tabular approaches like \xgb or \textsc{XGBoost} the industry’s go-to solutions for constructing predictive models on relational databases~\citep{grinsztajn2022tree}.

\subsection{R-GNNs}\label{subsec:graphbasedmodels}
With the growing popularity and success of GNNs~\citep{scarselli2008graph}, researchers have begun exploring their application to relational databases, particularly through the use of relational R-GNNs, which are designed to model heterogeneous relationships and leverage their ability to represent complex interactions in relational data. The most recent and prominent graph-based model in this context is \rdl \changed{(Relational Deep Learning)} \citep{fey2024position}. Given a seed time $t$, in \rdl, an heterogeneous directed graph is constructed in the following way:

\begin{definition}[Relational Graph up to time $t$]\label{def:graphrdl}
Given the relational database $\gR$ with set of entities $\gV$, the Relational Graph up to time $t$ is a heterogeneous directed graph $G(\gV_{\leq t}) = (\gV_{\leq t}, \gE_{\leq t},\phi, \psi)$, with $\phi$ as in \Cref{def:rdb} and:
\begin{enumerate}
    \item $\gV_{\leq t} = \{v \in \gV \mid t_v\leq t\}$ with $\gV$ as in \Cref{def:rdb};
    \item $\gE_{\leq t} = \{(v_1, v_2) \in \gV_{\leq t} \times \gV_{\leq t} \mid  k^{\phi(v_1)}_{v_2} = p_{v_1} \text{ and } t_{v_2}\leq t_{v_1}\}$ is the set of edges between entities, which captures connections between nodes based on primary-foreign key relationships;
    \item $\psi: \gE_{\leq t} \rightarrow \gL$, where $\gL \subseteq [N]\times[N]$ is the \textbf{edge type mapping} function that assigns each edge $(v_1, v_2) \in \gE_{\leq t}$ the pair of tables the belong to, i.e., $\psi(v_1, v_2) = (\phi(v_1), \phi(v_2))$.
\end{enumerate} 
\end{definition} 

\begin{figure}
    \centering
    \includegraphics[width=0.9\linewidth]{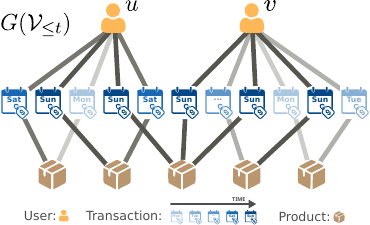}
    \caption{
    \changed{Example of relational graph $G(\gV_{\leq t})$ used in \rdl, where nodes represent users, products, and transactions up to time $t$ (here, "Sunday"). Each transaction is timestamped and linked to the corresponding user and product. The model predicts, for instance, the number of items a user will purchase the next day by aggregating interactions from previous time steps.}}
    \label{fig:graph_rdl}
\end{figure}

In \rdl, the relational graph up to time $t$ is used as input for a heterogeneous R-GNN~\citep{gilmer2017neural,fey2019fast,schlichtkrull2018modeling} to make predictions for the next time step. Specifically, the relational graph $G(\gV_{\leq t})$ enables inference for the next time interval by incorporating neighbors from previous time steps. \changed{For example, consider a snapshot of $G(\gV_{\leq t})$ shown in \Cref{fig:graph_rdl}, where the nodes represent users, transactions, and products. Each transaction is timestamped and associated with a specific day of the week. In this scenario, the prediction task may be to estimate the number of items a user will purchase on the following day. The graph $G(\gV_{\leq t})$ therefore includes all transaction nodes and their associated edges occurring up to time $t$, which in the figure corresponds to “Sunday”.} This cumulative aggregation across time steps enables temporal context but also results in a substantial growth of the graph size as $t$ increases, thereby increasing both computational complexity and inference time.

\paragraph{RDL Training and Inference Times.}
It is worth noting that while \rdl eliminates the need for manual feature engineering, many companies have already invested in developing robust feature engineering pipelines, which can provide a rich set of pre-existing features. 
Additionally, as a graph neural network model that needs to account for transactions in time, \rdl generally has higher computational requirements compared to tabular models like \xgb due to the graph size. This increased computational cost, even if compensated by somewhat better performance, can swing deployment decisions towards tabular models. 

\begin{table}[t]
    \caption{Inference time (seconds) and performance metrics (MAE and ROCAUC) for \xgb and \rdl. While both models demonstrate comparable performance, \rdl is significantly slower in inference time.}
    \begin{center}
    \resizebox{\linewidth}{!}{ 
    \begin{tabular}{lrr c rr}
        \toprule
         &  \multicolumn{2}{c}{Inf. time} & & \multicolumn{2}{c}{MAE}\\ 
         \cmidrule(l{2pt}r{2pt}){2-3}
         \cmidrule(l{2pt}r{2pt}){5-6}
         &  \xgb & \rdl & & \xgb & \rdl \\
        \midrule
        \texttt{f1} \textbf{driver-position} 
            & 0.04 & 2.34 & & 4.010 & 4.142\\
        \texttt{hm} \textbf{item-sales}
            & 0.13 & 10.52& & 0.038 & 0.056\\
        \texttt{event} \textbf{user-atten.}
            & 0.04 & 0.95 && 0.249 & 0.255\\
        \texttt{stack} \textbf{post-votes}
            & 0.79 & 35.23 && 0.068 & 0.065\\
        \texttt{amazon} \textbf{user-ltv}
            & 0.14 & 5.30 && 14.212 & 14.314\\
        \texttt{amazon} \textbf{item-ltv}
            & 0.04 & 5.48 & & 49.917 & 50.053 \\
        \midrule
         &  \multicolumn{2}{c}{Inf. time} & & \multicolumn{2}{c}{ROCAUC}\\ 
         \cmidrule(l{2pt}r{2pt}){2-3}
         \cmidrule(l{2pt}r{2pt}){5-6}
         &  \xgb & \rdl & & \xgb & \rdl \\
        \midrule
        \texttt{f1} \textbf{driver-dnf} 
            & 0.05 & 1.79 & &  70.52 & 71.08\\
        \texttt{f1} \textbf{driver-top3} 
            & 0.04 & 2.19 & &  82.77 & 80.30\\
        \texttt{hm} \textbf{user-churn} 
            & 0.3 & 3.63&  &  69.12 & 69.09\\
        \texttt{event} \textbf{user-ignore} 
            & 0.02 & 1.19 & &  82.62 & 77.82\\
        \texttt{event} \textbf{user-repeat} 
            & 0.04 & 2.29 & &  75.78 & 76.50\\
        \texttt{stack} \textbf{user-eng.} 
            & 0.14 & 24.01&  & 90.34 & 90.59\\
        \texttt{stack} \textbf{user-badge} 
            & 3.03 & 96.23&  & 86.34 & 88.54\\
        \texttt{amazon} \textbf{user-churn} 
            & 0.08 & 2.25& &  68.34 & 70.42\\
        \texttt{amazon} \textbf{item-churn} 
            & 0.08 & 2.24&  & 82.62 & 28.21\\
        \bottomrule
    \end{tabular}
    }
    \end{center}
    \label{tab:inference_time_MAE_xgb_rdl}
\end{table}

Table \ref{tab:inference_time_MAE_xgb_rdl} compares the inference times and predictive performance of \xgb (a tabular model) and \rdl across 15 datasets from the RelBench benchmark~\citep{robinson2024relbench}. The results show that while both models achieve comparable predictive quality (MAE for regression and ROCAUC for classification), \xgb significantly outperforms \rdl~\citep{robinson2024relbench} in inference speed.

This poses important research questions: What strategies can be employed to optimize the use of R-GNNs for relational databases, mitigating their computational overhead? Can we devise a hybrid approach that synergistically combines the strengths of R-GNNs and traditional tabular models, achieving a balance between predictive performance and computational efficiency?

To address these questions, we introduce \our, a new framework that harnesses the power of pre-trained tabular models to accelerate R-GNNs on relational databases.

\section{The \our Framework}\label{sec:method}

In this section we describe our proposed {\bf \our} framework.  \our  is a hybrid tabular-generalist modeling framework designed to integrate tabular data predictors (e.g., XGBoost, \xgb) with R-GNNs to capture both temporal and relational patterns. The core components of \our are the construction of a reduced relational graph that includes only a subset of interactions, rather than all prior ones, and a tabular model distillation approach.

\paragraph{(A) An Efficient Relational Graph:}
The most straightforward way to improve efficiency is by reducing the size of the graph used for training and inference. Instead of aggregating all interactions from all previous time steps, one could consider only the interactions that occurred immediately prior to the seed time of the task. By focusing solely on this restricted temporal window, the size of the relational graph is significantly reduced, enabling faster computation. Specifically, the idea is to apply an heterogeneous R-GNN to the relational graph defined as:
\begin{definition}[Snapshotted Relational Graph]
\label{def:TGt}
Given the \emph{relational database} $\gR$ with set of entities $\gV$, the Snapshotted Relational Graph is an \emph{heterogeneous directed graph} $G(\gV_{t}) = (\gV_{t}, \gE_{t},\phi, \psi)$, with $\phi$ as in \Cref{def:rdb}, $\psi$ as in \Cref{def:graphrdl} and:
\begin{enumerate}
    \item $\gV_{t} = \{v \in \gV \mid t_v= t\}$;
    \item $\gE_{t} = \{(v_1, v_2) \in \gV_{t} \times \gV_{t} \mid  k^{\phi(v_1)}_{v_2} = p_{v_1} \}$ is the set of edges between entities, capturing connections between nodes based on primary-foreign key relationships.
\end{enumerate} 
\end{definition}

\begin{figure}
    \centering
    \includegraphics[width=0.8\linewidth]{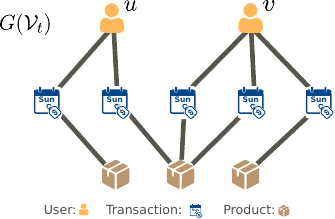}
    \caption{\changed{Example of Snapshotted Relational Graph $G(\gV_{t})$ used in \our, where nodes represent users, products, and transactions that occurred at time $t$ (here, “Sunday”). Unlike in \rdl, this graph includes only interactions from the current day, resulting in a significantly smaller and more efficient structure for training and inference.}}
    \label{fig:enter-label2}
\end{figure}

\changed{An illustrative example is shown in \Cref{fig:enter-label2}, where $G(\gV_{t})$ includes only the transactions from the current day, "Sunday", and their associated users and products}. 

In practice, when using the model in applications, the time $t$ can correspond to a specific day, week, or month. In our experiments, we chose a value for 
$t$ that ensures the window size matches that of the validation dataset, as detailed in \Cref{app:splittime}. Clearly, removing interactions that occurred further in the past entails losing valuable information necessary for studying temporal dynamics (see \Cref{sec:ablation} for results obtained when using the Snapshotted Relational Graph without incorporating any other temporal information). To address this, our proposed approach incorporates the temporal information by leveraging the knowledge learned by tabular methods trained on all historical data. This knowledge is integrated as additional node features in $G(\gV_{t})$. These features are generated through a knowledge distillation process from the tabular models, which we detail in the following. 
\changed{Furthermore, in \Cref{app:varing_windows_size}, we empirically assess the impact of varying the time window size. Results show that performance remains stable across different window lengths, confirming the robustness of our method. As expected, larger windows lead to an increase in inference time due to the larger amount of historical data being processed.}

\paragraph{(B) Tabular Model Distillation:}\label{par:dist}

Let $\sT^{(\leq t)}_i = \{\{\vv_v \in \sT_i : \forall v \in \gV_{\leq t}\}\}$ with table $\sT_i$ as in \Cref{def:rdb}.
Let $\texttt{FE}$ be a feature engineering process which assigns to each node $v\in \gV_t$ a new $d_f$-dimensional feature based on all the entities of all the tables up to time $t$, i.e., $\vf_{v,\leq t}=\texttt{FE}(v, \{\sT_i^{(\leq t)}\}_{i=1}^N)$. A feature-engineered tabular model (e.g., XGBoost, \xgb) is a model $\texttt{TB}$ trained on the set $\gD=\{(\vf_{v,\leq t},y_{v}):v\in \gV_{\leq t}\}$, where $y_{v}$ is the ground truth label associated to node $v\in \gV_{\leq t}$ i.e., $y_{v}\in \gC=\{c_1,\ldots,c_n\}$ for classification tasks or $y_{v}\in\sR$ for regression.
For all $v\in\gV_{\leq t}$, we define $\hat{y}^\texttt{TB}_{v}(\vf_{v,\leq t}):= \texttt{TB}(\vf_{v,\leq t})$.

Our model distillation is performed via multi-task learning over an MLP with two task heads as proposed in \citet{hinton2015distilling}.
The first task head is trained on the dataset $\gD=\{(\vf_{v,\leq t},y_{v}):v\in \gV_{\leq t}\}$. 
The second task head is trained on the predictions produced by \texttt{TB}, i.e., $\hat{\gD}=\{(\vf_{v,\leq t}, \hat{y}^\texttt{TB}_{v}(\vf_{v,\leq t})): v\in \gV_{\leq t}\}$.

Let $\hat{y}^{\text{MLP}}(\vf_{v,\leq t}):=\text{MLP}^{L}(\vf_{v,\leq t})$ be the softmax output of an $L$-layered MLP. The loss for the first task head is computed using the cross-entropy:

\begin{equation}
\mathcal{L}_{\text{hard}} = -\sum_{v \in \gV_{\leq t}} \sum_{c \in \mathcal{C}} {\bf 1}\{y_v = c\} \log (\hat{y}^{\text{MLP}}(\vf_{v,\leq t}))_c,
\end{equation}

where $(\hat{y}^{\text{MLP}}(\vf_{v,\leq t}))_c$ is the probability that the MLP assigns to class $c$ for entity $v$. 
For computing the loss for the second task head, we soften the output from the tabular model using a {softmax with a temperature parameter} $F\geq 1$,

\begin{equation}
p_c^{\texttt{TB},F}(\vf_{v,\leq t}) = \text{softmax}((\log \hat{y}^\texttt{TB}_{v}(\vf_{v,\leq t}))/F),
\end{equation}

where $F$ is the temperature parameter that controls the smoothness of the output distribution. 
The distillation loss is then computed as the cross-entropy between the soft labels provided by the tabular model and the soft labels generated by the MLP. This can be expressed as

\begin{equation}
\mathcal{L}_{\text{soft}} = -\sum_{v \in T} \sum_{c \in \mathcal{C}} p^{\texttt{TB},F}_c(\vf_{v,\leq t}) \log (\hat{y}^{\text{MLP}}(\vf_{v,\leq t}))_c.
\end{equation}

The total loss for the distillation process is a weighted combination of these two losses:

\begin{equation}
\mathcal{L}_{\text{total}} = \alpha \mathcal{L}_{\text{hard}} + (1 - \alpha) F^2 \mathcal{L}_{\text{soft}},
\end{equation}

where $\alpha $ is a hyperparameter that controls the trade-off between the ground truth learning and the distillation learning. In the case of regression tasks, we follow a similar procedure but replace the cross-entropy losses with an appropriate regression loss, i.e., the mean absolute error (MAE).

Once the MLP is trained, the embedding generated by the last hidden layer, $\textit{emb}_{v,\leq t}=\text{MLP}^{L-1}(\vf_{v,\leq t})$, with $\textit{emb}_{v,\leq t}\in \mathbb{R}^{d_e}$
contains the knowledge learned from the highly optimized features engineered tabular methods. These embeddings are then integrated as additional node features, that is, for every $v\in \gV_t$ with $t_v=t$:
\begin{equation}
    \vh_v^0 = \textit{emb}_{v,\leq t} \| \vx_v
\end{equation}
where $\vh^0_v \in \mathbb{R}^{d_e+d_i}$ is the initial representation of node $v$ for the R-GNN model. In principle, any existing tabular model from the literature can be used as the \texttt{TB} component; for our experiments we follow \citet{robinson2024relbench} and choose \xgb.
\changed{However, it is worth noting that \our is a general framework, and the tabular component is not limited to a specific model. In Appendix~\ref{app:alternative_tabular_models}, we demonstrate that \our maintains strong performance when \xgb is replaced with alternative tree-based models such as \textsc{CatBoost} and \textsc{XGBoost}, confirming the flexibility and robustness of the approach.}

It is worth addressing why knowledge distillation is necessary. One might argue for directly incorporating either: (1) the feature engineering into the graph nodes or (2) the pointwise predictions generated by \xgb. However, the feature engineering is specifically designed for tabular methods and does not generalize well to neural network models (see \Cref{app:our_no_time_with_FE}). Moreover, the pointwise prediction carries significantly less information compared to the embedding, which encapsulates the broader knowledge learned by the model (see \Cref{sec:ablation}). These considerations highlight the importance of the distillation approach adopted in \our.

In summary, \our follows these key steps: First, an efficient Snapshotted Relational Graph is constructed, containing only the most recent interactions relative to the time of prediction. Next, knowledge distillation is performed using \xgb trained on the complete historical dataset, generating an embedding that captures temporal dynamics. This embedding is incorporated as an additional feature for the nodes in the graph. Finally, a 
heterogeneous GraphSAGE model~\citep{fey2019fast, hamilton2017inductive}
is applied to the Snapshotted Relational Graph  to perform the desired tasks.

\begin{table*}[t]
\caption{Comparison of MAE and ROCAUC for \xgb, \rdl, and \our across Relbench tasks, averaged over five runs. Relative performance gains are highlighted in green, and relative speedups are highlighted in yellow. Refer to \Cref{app:res_with_std} for standard deviations. Often \our is faster than \xgb, since \our uses a distilled lightweight MLP with a lightweight R-GNN.}
\label{tab:comparison}
\setlength{\tabcolsep}{4pt}
\begin{center}
    \scriptsize
\begin{tabular}{lrrr>{\columncolor{mycustomcolor}}r>{\columncolor{mycustomcolor}}r>{\columncolor{mycustomcolor2}}r>{\columncolor{mycustomcolor2}}r}
\toprule
& \multicolumn{3}{c}{\textbf{MAE} ($\downarrow$)} & \multicolumn{2}{c}{\cellcolor{mycustomcolor}(\%) Gain w.r.t. ($\uparrow$)} & \multicolumn{2}{c}{\cellcolor{mycustomcolor2}Inference speedup w.r.t. ($\uparrow$)} \\
\cmidrule(l{2pt}r{2pt}){2-4}
\cmidrule(l{2pt}r{2pt}){5-6}
\cmidrule(l{2pt}r{2pt}){7-8}
& \xgb & \rdl & \our &  \xgb & \rdl & \xgb & \rdl \\
\midrule
\texttt{f1} \textbf{driver-position} & 4.010 & 4.142 & 3.861 & \textbf{+3.7} & \textbf{+6.8} & 1.0 & \textbf{58.5} \\
\texttt{hm} \textbf{item-sales} & 0.038 & 0.056 & 0.037 & \textbf{+2.6} & \textbf{+33.9} & \textbf{6.5} & \textbf{526.0} \\
\texttt{event} \textbf{user-atten.} & 0.249 & 0.255 & 0.238 & \textbf{+4.4} & \textbf{+6.7} & \textbf{2.0} & \textbf{47.5} \\
\texttt{stack} \textbf{post-votes} & 0.068 & 0.065 & 0.064 & \textbf{+4.7} & \textbf{+0.6} & \textbf{2.8} & \textbf{125.8} \\
\texttt{amazon} \textbf{user-ltv} & 14.212 & 14.314 & 13.587 & \textbf{+4.4} & \textbf{+5.1} & \textbf{1.6} & \textbf{58.9} \\
\texttt{amazon} \textbf{item-ltv} & 49.917 & 50.053 & 48.112 & \textbf{+3.6} & \textbf{+3.8} & 1.0 & \textbf{137.0} \\
\midrule
\multicolumn{4}{r}{\textbf{avg.}} &  \textbf{+3.9} & \textbf{+9.5} & \textbf{2.5}$\times$ faster & \textbf{159}$\times$ faster \\
\bottomrule
\\
& \multicolumn{3}{c}{\textbf{ROCAUC} ($\uparrow$)}& \multicolumn{2}{c}{\cellcolor{mycustomcolor}(\%) Gain w.r.t. ($\uparrow$)} & \multicolumn{2}{c}{\cellcolor{mycustomcolor2}Inference speedup w.r.t. ($\uparrow$)} \\
\cmidrule(l{2pt}r{2pt}){2-4}
\cmidrule(l{2pt}r{2pt}){5-6}
\cmidrule(l{2pt}r{2pt}){7-8}
& \xgb & \rdl & \our &  \xgb & \rdl & \xgb & \rdl  \\
\midrule
\texttt{f1} \textbf{driver-dnf} & 70.52 & 71.08 & 73.55 & \textbf{+4.3} & \textbf{+3.5} & \textbf{1.7} & \textbf{59.7} \\
\texttt{f1} \textbf{driver-top3} & 82.77 & 80.30 & 84.73 & \textbf{+2.4} & \textbf{+5.5} & 1.0 & \textbf{54.8} \\
\texttt{hm} \textbf{user-churn} & 69.12 & 69.09 & 68.93 & -0.3 & -0.2 & \textbf{15.0} & \textbf{181.5} \\
\texttt{event} \textbf{user-ignore} & 82.62 & 77.82 & 83.98 & \textbf{+1.6} & \textbf{+7.9} & 1.0 & \textbf{59.5} \\
\texttt{event} \textbf{user-repeat} & 75.78 & 76.50 & 77.77 & \textbf{+2.6} & \textbf{+1.7} & \textbf{2.0} & \textbf{114.5} \\
\texttt{stack} \textbf{user-eng.} & 90.34 & 90.59 & 89.02 & -1.5 & -1.7 & 0.5 & \textbf{80.0} \\
\texttt{stack} \textbf{user-badge} & 86.34 & 88.54 & 86.71 & \textbf{+0.4} & -2.1 & \textbf{9.5} & \textbf{300.7} \\
\texttt{amazon} \textbf{user-churn} & 68.34 & 70.42 & 69.87 & \textbf{+2.2} & -0.8 & 1.0 & \textbf{28.1} \\
\texttt{amazon} \textbf{item-churn} & 82.62 & 82.21 & 83.84 & \textbf{+1.5} & \textbf{+2.0} & \textbf{1.1} & \textbf{32.0} \\
\midrule
\multicolumn{4}{r}{\textbf{avg.}} & \textbf{+1.8} & \textbf{+1.7} & \textbf{3.6}$\times$ faster & \textbf{101.2}$\times$ faster \\
\bottomrule
\end{tabular}
\end{center}
\end{table*}

\section{Related Work}

Relational databases are integral to a wide range of applications, from e-commerce platforms~\citep{agrawal2001storage} and social media networks \citep{almabdy2018comparative} to banking systems~\citep{aditya2002banks} and healthcare services~\citep{park2014graph,kaur2015managing}. Although efforts to design deep learning architectures for tabular data have shown promising results~\citep{huang2020tabtransformer,arik2021tabnet,gorishniy2021revisiting,gorishniy2022embeddings,chen2023trompt,kim2024carte,holzmuller2024better,borisov2022deep,zhang2023GFS}, no deep-learning model has yet been demonstrated to clearly outperform tree-based methods on tabular data~\citep{shwartz2022tabular,mcelfresh2024neural,ye2024closer}. Recently, a new transformer-based deep learning method for tabular data, TabPFN, was shown to achieve competitive results on small to medium-sized datasets~\citep{hollmann2025accurate}. However, its scalability is limited due to its memory requirements and inference complexity, making it not suited for our relational databases.

Among the deep learning models proposed for relational data, GNNs have gained attention~\citep{schlichtkrull2018modeling,ioannidis2019recurrent,cvitkovic2020supervised,zhang2020relational,zahradnik2023deep,ferrini2024meta,ferrini2025a}. Recently, \citet{fey2024position,robinson2024relbench} proposed a general end-to-end learnable framework for solving tasks on relational data that incorporates a temporal dimension. However, this method also struggles to outperform classic tabular methods when equipped with feature engineering and is significantly less efficient than them. We propose combining tabular methods, such as \xgb, with GNNs to achieve better performance and efficiency.

Other works have explored the combination of GNNs with boosting methods~\citep{ivanov2021boost,sunadagcn,shi2021boosting,zheng2021adaboosting,deng2021xgraphboost,yan2024xgboost,tang2024xgnn,fan2023spatial,oono2020optimization}. However, they focus on improving standard GNNs for graph datasets that are not derived from relational databases, and therefore lack temporal and heterogeneous components. Moreover, while these methods often aim to replace tabular data models with GNNs within a boosting setup, we instead incorporate pretrained tabular models as a dedicated component for modeling temporality. This is complemented by a R-GNN to capture the structural relationships in the data. 

\section{Results}
\label{sec:results}
The goal of our experiments is to show that \our significantly reduces both training and inference times compared to \rdl while also improving accuracy. We considered RelBench~\citep{robinson2024relbench}, a comprehensive benchmark for relational databases. RelBench was chosen not only for its comprehensive coverage and diversity of relational datasets but also because, for most datasets, it provides pre-designed high-quality feature engineering, which is critical for evaluating \our which integrate tabular and graph-based approaches. 
\changed{We focused on \textbf{all RelBench datasets} where expert-designed feature engineering was available for node-level tasks. This is not a limitation of our method but a deliberate choice to ensure fair comparisons: by relying on existing features rather than engineering new ones ourselves, we avoid the risk of overfitting or unintentionally tailoring features to our method. This setup ensures that \our is evaluated using the same feature inputs as the baselines, preserving the fairness and integrity of the comparison.}
A detailed description of the RelBench datasets is provided in \Cref{app:dataset}, while detailed model configurations and training procedures are provided in \Cref{app:Exp_setup_and_rep}. The code to reproduce the experimental results is publicly available at \url{https://github.com/AntonioLonga/LightRDL}.

\subsection{\our Configuration and Baselines}

\our is configured using two models: (a) the pretrained feature-engineered \xgb model from RelBench~\citep{robinson2024relbench}; and (b) a R-GNN (\rgcn) that is a HeteroGraphSAGE~\citep{hamilton2017inductive}, identical to the R-GNN used in the \rdl model~\citep{robinson2024relbench}, but applied to the Snapshotted Relational Graph instead of the original relational graph. We evaluate \our against two baselines which are the strongest-performing methods in \citet{robinson2024relbench}: the pretrained feature-engineered \xgb model from RelBench and the \rdl model. The architectural details of the different parts of \our are provided in \Cref{app:Exp_setup_and_rep}, and the specifics of the distillation process can be found in \Cref{app:distillation_details}.

\changed{We focus exclusively on the tasks in RelBench that include a feature engineering strategy. This is not a limitation of our method but a deliberate choice to ensure fair comparisons: by relying on existing, expert-crafted features rather than engineering new ones ourselves, we avoid the risk of overfitting or unintentionally tailoring features to our approach. This setup ensures that our method is evaluated on the same footing as prior baselines, with a consistent and unbiased feature space.}

\begin{figure*}
    \centering
    \includegraphics[width=\linewidth]{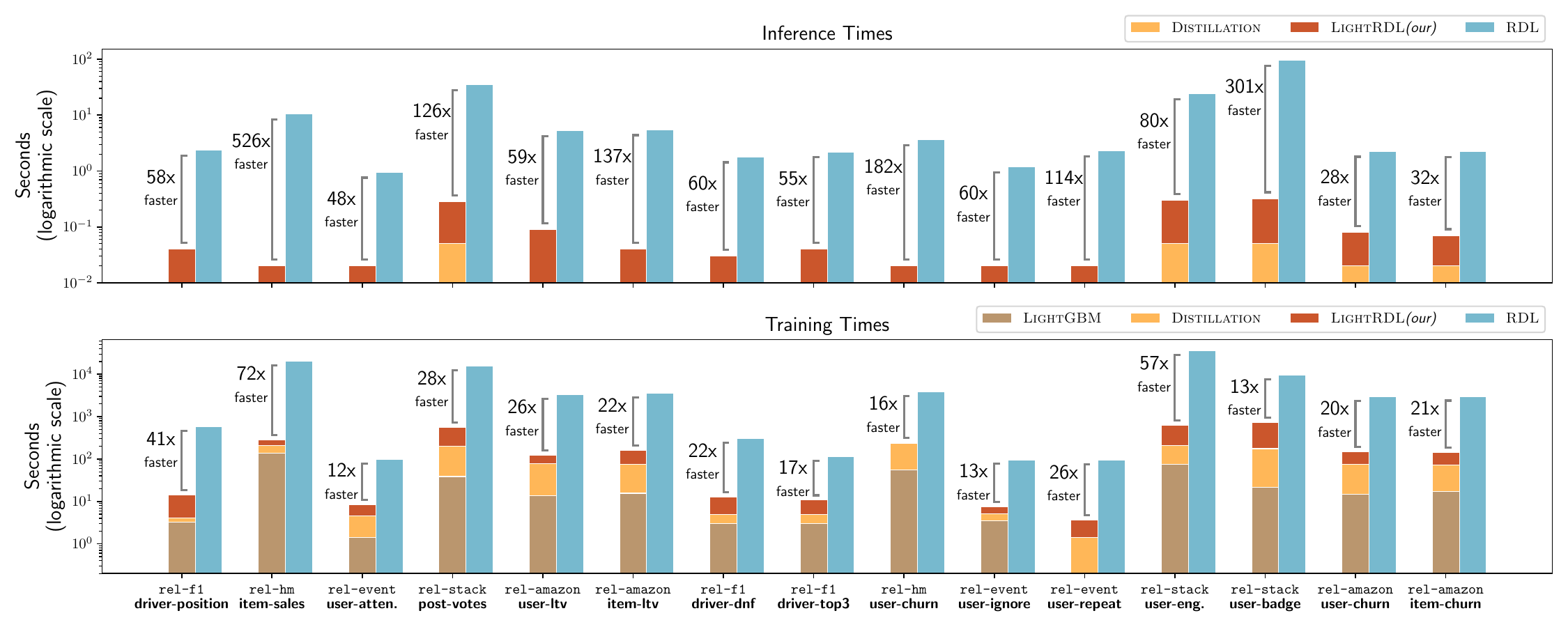}
    \vspace{-19pt}
    \caption{Comparison of inference and training times for \our and \rdl, across tasks in the Relbench dataset (logarithmic scale); \our achieves speedups in inference, ranging from 28$\times$ to 526$\times$ compared to \rdl. Similarly, \our is up to 72$\times$ faster than \rdl in training. Total inference time for \our is calculated summing up the inference time of the GNN model (red bar) and inference time of the distilled model (yellow bar). Total training time is calculated summing up the training time of the GNN model (red bar), the time required to train the distilled model (yellow bar) and the training time of \xgb (brown bar).}
    \label{fig:time}
\end{figure*}

\subsection{Experimental Results}

\paragraph{Regression and Classification Performance.}
~\Cref{tab:comparison} shows that, for the regression tasks, \our consistently outperforms both \xgb and \rdl, achieving an average relative gain of +9.5\% over \rdl and +3.9\% over \xgb, reaching a maximum gain of +33.9\%. \changed{For a broader comparison with additional baselines, including linear and neural tabular models, refer to Appendix~\ref{app:regression_baselines}.} For the classification tasks, \our shows an average improvement of +1.7\% over \rdl and +1.8\% over \xgb. In classification, the improvement is not as high as in regression, and in some tasks, such as \textbf{user-badge}, \our shows worse performance compared to \rdl (-2\%). However, this is balanced by a significant increase in efficiency: \our is 13$\times$ faster in training and 301$\times$ faster in inference on this dataset (see \Cref{fig:time}), which makes our method highly competitive. A more detailed analysis of the runtime performance is presented in the next paragraph.

\paragraph{Inference Time.}
\our achieves a substantial reduction in inference times compared to \rdl across all datasets and tasks, with speedups ranging from 28$\times$ to 526$\times$ (\Cref{fig:time}). 
For a fair comparison of inference times, we considered the total inference time of our model as the sum of the inference time of our model and the time spent on distillation (yellow bar + red bar in the figure). The speedups with respect to \rdl largely compensate for the occasional limited drop in classification performance (\Cref{tab:comparison}). Indeed, the largest decrease in classification performance is 2.1\% in the \textbf{user-badge} task, where \our has a 301-fold increase in inference speed over \rdl. This substantial improvement makes \our particularly well-suited for relational database ML applications, which demand low-latency and low-computational overhead. As a result, \our is not only preferred for scenarios requiring rapid response times but is also highly practical for deployment in industrial settings where pretrained tabular models already exist, and computational resources and inference latency are critical constraints.
The inference time of \our is lower or comparable also to that of \xgb (see \Cref{app:inftime}). \\
These experiments seek to show that organizations with already existing high performing tabular models can leverage \our to enhance their performance and seamlessly integrate new features and unstructured data (e.g., images, text embeddings)
into their machine learning pipelines.

\paragraph{Training Time.}
Achieving state-of-the-art performance with deep neural network models often requires extensive fine-tuning. Consequently, training time becomes a critical factor, as the fine-tuning process can be expensive in terms of both time and computational resources. \Cref{fig:time} reports the training times. 
To calculate the complete training time of \our, we added the training times of \xgb with the distillation time and the training time of the R-GNN model. The reported training times refer to a single seed using the model's best configuration. Details on the hyperparameter search are provided in \Cref{app:Exp_setup_and_rep}; for a detailed breakdown of the training time per epoch and the mini-batch strategy, see~\Cref{app:train_time_epoch}. \our model is consistently faster than \rdl, with \our achieving 13$\times$ and 72$\times$ faster training. This dramatic speedup is due primarily to two factors: (i) the effect of the number of timestamps of the tasks, which does not affect \our and significantly affects \rdl: unlike \rdl, where the training graph $G(\gV_{\leq t})$ is constructed using entities and relations of all the time up to $t$~\citep{robinson2024relbench}, \our graph $G(\gV_t)$ only uses the information at the timestamp $t$ before the inference, so the size of its graph is independent of the number of snapshots; (ii) the much smaller number of model parameters in \our (see Appendix~\ref{app:param}). Indeed, \rdl uses a ResNet tabular model~\cite{hu2024pytorch,gorishniy2021revisiting} to encode the features. This model is trained end-to-end, significantly increasing the computational burden and resulting in up to 5,942,785 parameters (details in \Cref{app:param}). 

\Cref{app:lessParam} reports other attempts to speed-up \rdl by reducing its number of parameters, which yielded only minor speed improvements while drastically impairing \rdl's predictive quality. In contrast, \our achieves greater efficiency by employing \xgb to generate features. Unlike a simple feature encoder, \xgb plays a more significant role as it is trained on all historical data, embedding not only the static information of each node but also its temporal context. Additionally, since \xgb is not trained end-to-end, it ensures faster training while providing richer and more comprehensive information.

\paragraph{Time for Constructing Relational Graph from Relational Database.}

\our is also efficient in the graph construction phase. \Cref{tab:prepocessing_time} reports the time taken by \rdl and \our to build the relational graph up to time $t$ (\Cref{def:graphrdl}) and the Snapshotted Relational Graph (\Cref{def:TGt}) from the relational database, respectively. On average, \our is twice as fast as \rdl in this process. This improvement is due to the fact that \our considers only the table entries corresponding to a specific time, rather than including all prior entries.

\begin{table}[t]
    \caption{Time to construct relational graph (seconds). This time is {\bf not counted} as training/test time in other results. On average, \rdl takes twice as long to build the graph compared to \our.}
    \begin{center}
    
    \scriptsize
    \begin{tabular}{p{0.3mm}lrr}
        \toprule
        && \rdl & {\our} \\
        \midrule 
        \multirow{6}{*}{\rotatebox[origin=c]{90}{Regression}}
        & \texttt{rel-f1} \textbf{driver-position} 
            & 8 & 31 \\
        & \texttt{rel-hm} \textbf{item-sales} 
            & 1126 & 48 \\
        & \texttt{rel-event} \textbf{user-atten.} 
            & 68 & 19 \\
        & \texttt{rel-stack} \textbf{post-votes} 
            & 1033 & 530 \\
        & \texttt{rel-amazon} \textbf{user-ltv} 
            & 486 & 324 \\
        & \texttt{rel-amazon} \textbf{item-ltv} 
            & 501 & 351 \\
        \midrule 
        
        \multirow{9}{*}{\rotatebox[origin=c]{90}{Classification}}
        & \texttt{rel-f1} \textbf{driver-dnf}  & 3 &  11 \\
        & \texttt{rel-f1} \textbf{driver-top3}  & 2&   15\\
        & \texttt{rel-hm} \textbf{user-churn}  & 750&   51\\
        & \texttt{rel-event} \textbf{user-ignore}  & 407&  20\\
        & \texttt{rel-event} \textbf{user-repeat}  & 54&  8\\
        & \texttt{rel-stack} \textbf{user-eng.} & 1079&  571 \\
        & \texttt{rel-stack} \textbf{user-badge}  & 1052&  595\\
        & \texttt{rel-amazon} \textbf{user-churn}   &180 & 258 \\
        & \texttt{rel-amazon} \textbf{item-churn}  & 210  & 341  \\   
        \midrule
        \multicolumn{2}{r}{\textbf{avg.}} & 464 & 212\\
        \bottomrule
    \end{tabular}
    \end{center}
    \label{tab:prepocessing_time}
    \vspace{-20pt}
\end{table}

\subsection{Ablation study}\label{sec:ablation}
We conducted an ablation study to address two key questions: (i) is temporal information necessary for relational database tasks? (ii) Is the distillation of boosted tree models truly essential?
To answer these questions, we compare the performance of \our against two baseline models: \our \textsc{w/o time}, which is a HeteroGraphSAGE without any temporal information \changed{(and is conceptually similar to \rdl with \texttt{temporal\_strategy = last})}, and \our \textsc{w P.}, which incorporates temporal information but without distillation, instead directly integrating the row predictions produced by \xgb as additional node features.
The results provide clear answers to both questions. First, temporal modeling shows to be critical for predictive tasks on relational databases. As shown in \Cref{tab:ablation}, \our \textsc{w/o time} consistently underperforms when compared to the models that incorporate temporal information (column 1 vs. columns 2 and 3).

Second, the distillation process is also essential. \our significantly outperforms \our \textsc{w P.}, demonstrating that embedding the distilled knowledge offers a more informative and effective way to enrich node features than using raw predictions.

\begin{table}[t]
    \caption{The ablation study shows that (i) temporal modeling is critical for predictive tasks on relational databases, see \our w/o time v.s.\ \our and (ii) the distillation process leads to better performance, see \our with \xgb prediction (w.P.)\ v.s.\ \our. }
    \begin{center}
        \setlength{\tabcolsep}{2.4pt}
        \scriptsize
    \begin{tabular}{p{1.2mm}p{1.2mm}lrrr}
        \toprule
        &&    &  \our   & \our & \our \\
        &&   &   \textsc{w/o time} & \textsc{w.P.} & \\
        \midrule
        \multirow{6}{*}{\centering \rotatebox[origin=c]{90}{Regression} }&
        \multirow{6}{*}{\centering \rotatebox[origin=c]{90}{(MAE)} }&
        \texttt{rel-f1} \textbf{driver-position} &   \val{5.604} & \val{3.941} & \valf{3.861} \\
        &&\texttt{rel-hm} \textbf{item-sales} &   \val{0.055}      & \val{0.038}  & \valf{0.037} \\
        &&\texttt{rel-event} \textbf{user-atten.} & \val{0.261}  & \val{0.242} & \valf{0.238} \\
        &&\texttt{rel-stack} \textbf{post-votes} &   \val{0.123}  & \val{0.068} & \valf{0.064}  \\
        &&\texttt{rel-amazon} \textbf{user-ltv} &  \val{16.881}  & \val{14.088} & \valf{13.587}  \\
        &&\texttt{rel-amazon} \textbf{item-ltv} &   \val{57.323}  & \val{49.314} & \valf{48.112}  \\
        \midrule
        \multirow{9}{*}{\centering \rotatebox[origin=c]{90}{Classification} }&
        \multirow{9}{*}{\centering \rotatebox[origin=c]{90}{(ROCAUC)} }&
        \texttt{rel-f1}  \textbf{driver-dnf} &  \val{68.80} & \val{70.79} & \valf{73.55} \\
        &&   \texttt{rel-f1}   \textbf{driver-top3} & \val{77.01}  & \val{78.77} & \valf{84.73}\\
        &&  \texttt{rel-hm}  \textbf{user-churn}  & \val{56.07}      & \valf{69.38}  & \val{68.93} \\
        &&  \texttt{rel-event}  \textbf{user-ignore} &   \val{80.60}  & \val{78.76} & \valf{83.98} \\
        &&  \texttt{rel-event}  \textbf{user-repeat} &   \val{69.01}  & \val{73.37} & \valf{77.77} \\
        &&  \texttt{rel-stack} \textbf{user-eng.}  &   \val{78.58}  & \valf{89.94} & \val{89.02}  \\
        &&   \texttt{rel-stack} \textbf{user-badge} &   \val{81.01} & \val{84.12} &  \valf{86.71} \\
        &&  \texttt{rel-amazon}  \textbf{user-churn} &   \val{67.58}  & \val{69.18} & \valf{69.87}  \\
        && \texttt{rel-amazon}  \textbf{item-churn}&   \val{79.58}  & \val{83.37} & \valf{83.84}  \\
        \bottomrule
    \end{tabular}
    \end{center}
    \label{tab:ablation}
    \vspace{-20pt}
\end{table}

\section{Conclusion}
In this work, we introduced \our, a hybrid framework that combines the efficiency and temporal modeling capabilities of pretrained tabular models with the relational expressiveness of graph-based methods. By leveraging knowledge distillation and integrating highly optimized tabular features, \our bridges the gap between these approaches, achieving superior predictive performance while achieving good computational efficiency.

Our experimental results demonstrate that \our outperforms state-of-the-art graph-based models like \rdl in predictive accuracy, with up to $33\%$ performance improvement, and delivers substantial gains in inference speed, being $526\times$ faster. Compared to tabular models like \xgb, \our consistently achieves better predictive accuracy while maintaining reasonable efficiency.

These results highlight the practical benefits of integrating tabular and graph-based models, providing a scalable and efficient solution for predictive tasks on relational databases.

\section*{Acknowledgments}
BR acknowledges the National Science Foundation (NSF) awards, CCF-1918483, CAREER IIS-1943364 and CNS-2212160, Amazon Research Award, AnalytiXIN, and the Wabash Heartland Innovation Network (WHIN), Ford, NVidia, CISCO, and Amazon. Computing infrastructure was supported in part by CNS-1925001 (CloudBank). This work was supported in part by AMD under the AMD HPC Fund program.
 \clearpage
 \newpage

\bibliography{example_paper}
\bibliographystyle{icml2025}

\appendix
\onecolumn

\section{Datasets}\label{app:dataset}
The Relational Deep Learning Benchmark (RelBench) is a collection of large-scale, real-world benchmark datasets designed for machine learning on relational databases. We consider \emph{all} the datasets in RelBench for which feature engineering has been performed, as listed below.

\paragraph{\texttt{rel-event}}
originates from the Hangtime mobile app, which tracks users' social plans and interactions with friends. The dataset contains anonymized data on user actions, event metadata, and demographic information, as well as users’ social connections, allowing for an analysis of how these relations influence behavior. No personally identifiable information is included in the dataset. The entity predictive tasks on this database are:
\begin{itemize}
    \item \textbf{user-atten.}: Predict how many events each user will respond "yes" or "maybe" to within the next seven days.
    \item \textbf{user-repeat}: Determine whether a user will attend another event (by responding "yes" or "maybe") within the next seven days, given they have attended an event in the last 14 days.
    \item \textbf{user-ignore}: Predict whether a user will ignore more than two event invitations within the next seven days.
\end{itemize}

\paragraph{\texttt{rel-f1}}
comprises historical data and statistics from Formula 1 racing, covering the period from 1950 to the present. It includes comprehensive information on key stakeholders, such as drivers, constructors, engine manufacturers, and tire manufacturers. The dataset highlights geographical details of circuits, along with detailed historical season data, including race results, practice sessions, qualifying positions, sprints, and pit stops. The entity predictive tasks on this database are:
\begin{itemize}
    \item \textbf{driver-position}: Forecast the average finishing position of each driver across all races in the upcoming two months.
    \item \textbf{driver-dnf}: Predict whether a driver will fail to complete a race (DNF - Did Not Finish) within the next month.
    \item \textbf{driver-top3}: Determine whether a driver will qualify within the top 3 positions in a race over the next month.
\end{itemize}

\paragraph{\texttt{rel-hm}}
comprises extensive customer and product data from the company’s online shopping platform of H\&M. It includes detailed customer purchase histories and a wide range of metadata, covering everything from customer demographics to comprehensive product information. This dataset enables a deep analysis of shopping behavior across a broad network of brands and stores. The entity predictive tasks on this database are:
\begin{itemize}
    \item \textbf{user-churn}: Predict customer churn (i.e., no transactions) within the next week.
    \item \textbf{item-sales}: Estimate the total sales for a product (summed over the associated transactions) during the next week.
\end{itemize}

\begin{table}[t]
\caption{Tasks details.}\label{tab:datadet}
\begin{center}
\scriptsize
\addtolength{\tabcolsep}{-0.45em}
\begin{tabular}{@{}l l r r r r r@{}}
\toprule
\textbf{Dataset} and \textbf{Task} name            & Task type      & \#Rows Train  & \#Rows Val & \#Rows Test  & \#Unique Entities & \%train/test Entity Overlap \\ \midrule
\texttt{rel-event}  \textbf{user-atten.}   & entity-reg      &  19,261   & 2,014 & 2,006   & 9,694 & 14.6 \\
\texttt{rel-event}  \textbf{user-repeat}       & entity-cls      & 3,842    & 268 & 246   & 1,514 & 11.5 \\
\texttt{rel-event}  \textbf{user-ignore}       & entity-cls      & 19,239   & 4,185 & 4,010   & 9,0799 & 21.1 \\ \midrule
\texttt{rel-f1}  \textbf{driver-dnf}        & entity-cls      & 11,411   & 566 & 702   & 821 & 50.0 \\
\texttt{rel-f1}  \textbf{driver-top3}       & entity-cls      & 1,353    & 588 & 726   & 134 & 50.0 \\
\texttt{rel-f1}   \textbf{driver-position}   & entity-reg      & 7,453    & 499 & 760   & 826 & 44.6 \\ \midrule
\texttt{rel-hm}   \textbf{user-churn}        & entity-cls      & 3,871,410 & 76,556 & 74,575   & 1,002,984 & 89.7 \\
\texttt{rel-hm}   \textbf{item-sales}        & entity-reg      & 5,488,184 & 105,542 & 105,542   & 1,005,542 & 100.0 \\ \midrule
\texttt{rel-stack}  \textbf{user-engagement}   & entity-cls      & 1,360,850 & 85,838 & 88,137   & 88,137 & 97.4 \\
\texttt{rel-stack}  \textbf{user-badge}        & entity-cls      & 3,386,276 & 247,398 & 255,360   & 255,360 & 96.9 \\
\texttt{rel-stack}  \textbf{post-votes}        & entity-reg      & 2,453,921 & 156,216 & 160,903   & 160,903 & 97.1 \\ \midrule
\texttt{rel-amazon}  \textbf{user-churn}      & entity-cls      & 4,732,555 &409,792 & 351,885   & 1,585,983 & 88.0 \\
\texttt{rel-amazon} \textbf{item-churn}        & entity-cls      & 2,559,264 & 177,689 & 166,842   & 416,352 & 93.1 \\
\texttt{rel-amazon} \textbf{user-ltv}         & entity-reg      & 4,732,555 & 409,792 & 351,885   & 1,585,983 & 88.0 \\
\texttt{rel-amazon} \textbf{item-ltv}          & entity-reg      & 2,707,679 & 166,978 & 178,334   & 427,537 & 93.5 \\ \bottomrule
\end{tabular}

\end{center}

\end{table}

\paragraph{\texttt{rel-stack}}
captures detailed interactions from the network of question-and-answer websites Stack Exchange. It includes comprehensive records of user activity, such as biographies, posts, comments, edit histories, voting patterns, and links between related posts. The reputation system within Stack Exchange enables self-moderation of the community. In our experiments, we use data from the stats-exchange site. The entity predictive tasks on this database are:
\begin{itemize}
    \item \textbf{user-engagement}: Predict whether a user will engage (e.g., through votes, posts, or comments) within the next three months.
    \item \textbf{post-votes}: Forecast how many votes a user’s post will receive over the next three months.
    \item \textbf{user-badge}: Predict if a user will be awarded a new badge during the next three months.
\end{itemize}

\paragraph{\texttt{rel-amazon}}
The Amazon E-commerce database documents products, users, and reviews from Amazon's platform, providing comprehensive details about products and their associated reviews. Each product entry includes its price and category, while reviews capture the overall rating, whether the reviewer purchased the product, and the review text. For our analysis, we focus specifically on a subset of book-related products.

\begin{itemize}
    \item \textbf{user-churn}: Predict whether a user will refrain from reviewing any products in the next three months (1 for no reviews, 0 otherwise).
    \item \textbf{item-churn}: Determine if a product will receive no reviews in the next three months (1 for no reviews, 0 otherwise).
    \item \textbf{user-ltv}: Predict the value of the total products a user purchases and reviews over the next three months.
    \item \textbf{item-ltv}: Predict the value of the total purchases and reviews a product receives in the next three months.
\end{itemize}

Further details regarding the tasks are provided in \Cref{tab:datadet}.

\section{Additional Regression Baselines}
\label{app:regression_baselines}
\changed{To further strengthen our evaluation, we include additional baselines for all regression tasks in RelBench. Performance is reported in terms of mean absolute error (MAE), where lower values indicate better performance. The following baselines were considered:}

\changed{\begin{itemize}
    \item \textbf{LinReg}: A simple linear regressor trained on the feature-engineered inputs.
    \item \textbf{MLP}: A multilayer perceptron with two hidden layers.
    \item \textbf{LGBM (w/o FE)}: \xgb trained on raw relational features (no feature engineering).
    \item \textbf{LGBM}: \xgb with the original feature-engineered inputs from RelBench.
    \item \textbf{XGB}: The pretrained XGBoost model from RelBench.
    \item \textbf{RDL (w.~P.)}: RDL using raw \xgb predictions as node features.
    \item \textbf{RDL (w.~D.)}: RDL using distilled embeddings from \xgb.
    \item \textbf{RDL}: Original RDL model from \citet{robinson2024relbench}.
    \item \textbf{LightRDL}: Our proposed method.
\end{itemize}}

\changed{LightRDL achieves the best performance across all tasks except \texttt{driver-pos.}, where RELGNN and Llama-based methods perform slightly better.}

\begin{table}[t]
\scriptsize
\caption{\changed{Comparison of MAE scores (lower is better) for regression tasks. \our outperforms all baselines in 5 out of 6 tasks.}}
\begin{center}
\begin{tabular}{lrrrrrrrrr}
\toprule

\textbf{Dataset} and \textbf{Task} & \textsc{Linear} & \textsc{MLP} & \xgb & \xgb & \textsc{XGB} & \rdl & \rdl & \rdl & \our \\
 & \textsc{Regressor} &  & (w/o FE) & &  & (w.P.) & (w.D.) &  & \\

\midrule
\texttt{rel-f1} \textbf{driver-position}   & 4.690 & 4.350 & 4.170 & 4.010 & 4.850 & 3.990 & 4.120 & 4.140 & \textbf{3.860} \\
\texttt{rel-hm} \textbf{item-sales}    & 0.709 & 0.076 & 0.076 & 0.038 & 0.038 & 0.050 & 0.052 & 0.056 & \textbf{0.037} \\
\texttt{rel-event} \textbf{user-atten}   & 0.440 & 0.260 & 0.260 & 0.240 & 0.240 & 0.240 & 0.240 & 0.250 & \textbf{0.230} \\
\texttt{rel-stack} \textbf{post-votes}    & 0.268 & 0.069 & 0.068 & 0.068 & 0.071 & 0.065 & 0.065 & 0.065 & \textbf{0.064} \\
\texttt{rel-amazon} \textbf{user-ltv}      & 16.800 & 16.800 & 60.600 & 14.200 & 14.100 & 14.200 & 13.900 & 14.300 & \textbf{13.600} \\
\texttt{rel-amazon} \textbf{item-ltv}      & 60.800 & 60.700 & 16.800 & 49.900 & 50.100 & 49.200 & 48.700 & 50.100 & \textbf{48.100} \\
\bottomrule
\end{tabular}
\label{tab:regression_baselines_clean}
\end{center}
\end{table}

\section{\changed{Generality of \our: Alternative Tabular Models}}
\label{app:alternative_tabular_models}

\changed{\our is designed as a general framework that allows flexibility in the choice of both its tabular and graph-based components. While our main experiments adopt \xgb as the default tabular model—consistent with the RelBench benchmark—we also evaluate the framework using alternative tree-based models.}

\changed{Specifically, we conducted additional experiments on the \texttt{rel-f1} tasks by replacing \xgb with \textsc{CatBoost} and \textsc{XGBoost} in the distillation pipeline. The results, reported in \Cref{tab:tabular_swap}, demonstrate that \our maintains high performance regardless of the specific tabular model used. Performance variations across the different models are minor, which supports the claim that the core mechanism of \our is robust and model-agnostic with respect to the tabular backend.}

\begin{table}[t]
\scriptsize
\caption{\changed{Evaluation of \our using different tree-based models in the distillation pipeline. MAE is reported for \texttt{driver-position}, while ROCAUC is used for the classification tasks.}}
\begin{center}
\begin{tabular}{lrrr}
\toprule
\textbf{Task} & \textbf{\our} & \textbf{CatBoost} & \textbf{XGBoost} \\
\midrule
\texttt{driver-position} (MAE) & 3.861 & 3.846 & 3.895 \\
\texttt{driver-dnf} (ROCAUC)   & 73.55 & 72.56 & 73.93 \\
\texttt{driver-top3} (ROCAUC)  & 84.73 & 83.41 & 83.87 \\
\bottomrule
\end{tabular}
\label{tab:tabular_swap}
\end{center}
\end{table}

\section{Parameters}\label{app:param}
\Cref{tab:number_of_param} presents the number of parameters of \our and \rdl across the different tasks. On average, \our uses 73.09\% fewer parameters than \rdl.

\begin{table}[t]
    \caption{Number of parameters. On average, \our uses 73.09\% fewer parameters than \rdl.}\label{tab:number_of_param}
    \begin{center}
    \scriptsize
    \begin{tabular}{llrr}
        \toprule
        & \textbf{Dataset} and \textbf{Task} & \rdl   & \our \\
        \midrule
        \multirow{9}{*}{\centering \rotatebox[origin=c]{90}{Classification} }
        &  \texttt{rel-f1}  \textbf{driver-dnf}  &  5,073,793 & 271,803 \\
        &  \texttt{rel-f1}  \textbf{driver-top3}  & 5,073,793  & 648,395\\
        &  \texttt{rel-hm}  \textbf{user-churn}  & 2,178,945  & 22,204 \\
        &  \texttt{rel-event}  \textbf{user-ignore}   & 5,942,785  & 1,231,871\\
        &  \texttt{rel-event}  \textbf{user-repeat}   & 5,942,785  & 467,583 \\
        &  \texttt{rel-stack}  \textbf{user-engagement}  & 4,322,177  &  2,847,127\\
        &  \texttt{rel-stack}  \textbf{user-badge} &   4,322,177 & 3,586,454 \\
        &  \texttt{rel-amazon}  \textbf{user-churn} &  5,129,348  &  1,622,173\\           &  \texttt{rel-amazon}  \textbf{item-churn} &  5,129,348 & 1,730,512\\
        \midrule
        \multirow{6}{*}{\centering \rotatebox[origin=c]{90}{Regression} }
        &  \texttt{rel-f1}  \textbf{driver-position}   & 5,073,793 & 1,372,395  \\
        &  \texttt{rel-hm}  \textbf{item-sales} &   2,178,945    & 73,409 \\
        &  \texttt{rel-event}  \textbf{user-atten.}   & 5,942,785  & 1,518,271 \\
        &  \texttt{rel-stack}  \textbf{post-votes}   & 4,322,177  & 1,987,955 \\
        &  \texttt{rel-amazon}  \textbf{user-ltv}   & 5,129,348  &   925,571\\
        &   \texttt{rel-amazon}  \textbf{item-ltv}   & 5,129,348 & 1,113,729\\
        \bottomrule
    \end{tabular}
    \end{center}
\end{table}

\section{\rdl with Less Parameters}\label{app:lessParam}
In this section, we compare \our with a reduced-parameter version of \rdl, referred to as \rdl small. The number of parameters in \rdl small was reduced to match the scale of \our. Tables \ref{tab:rdl_less_param_class} and \ref{tab:rdl_less_param_reg} present the metrics (ROCAUC/MAE), training time, and inference time. The results demonstrate that reducing the parameters in \rdl leads to slightly faster training and inference times but comes at the cost of diminished performance.
However, even with this reduction, \rdl small remains significantly slower than \our in both training and inference, highlighting the efficiency advantage of \our.

\begin{table}[t]
    \caption{\rdl with reduced parameters (21,449 parameters) for the classification task shows a decrease in performance. Moreover, reducing the parameter does not offer the speedup of \our, indicating that the speedup of \our comes mainly from the Snapshotted Relational Graph not from the reduced number of parameters.}
    \begin{center}
    \scriptsize
    {
        \begin{tabular}{lcccccc}        
        \toprule    
        \textbf{Dataset} and \textbf{Task} & \multicolumn{2}{c}{ROCAUC ($\uparrow$)}  & \multicolumn{2}{c}{Training time (seconds)} & \multicolumn{2}{c}{Inference time (seconds)}\\
        \cmidrule(l{2pt}r{2pt}){2-3}
        \cmidrule(l{2pt}r{2pt}){4-5}
        \cmidrule(l{2pt}r{2pt}){6-7} 
        & \rdl small & \our &  \rdl small & \our &  \rdl small & \our  \\
        \midrule
        \texttt{rel-hm} \textbf{user-churn} & 64.28 & \textbf{68.93} & 1635 & \textbf{13} & 1.39 &\textbf{0.33}\\
        \bottomrule
        \end{tabular}
    }
    \end{center}
    \label{tab:rdl_less_param_class}
\end{table}

\begin{table}[t]
    \caption{\rdl with reduced parameters (21,449 parameters) for the regression task shows a decrease in performance. Moreover, reducing the parameter does not offer the speedup of \our, indicating that the speedup of \our comes mainly from the Snapshotted Relational Graph not from the reduced number of parameters.}
    \begin{center}
    \scriptsize
    {
        \begin{tabular}{lcccccc}        
        \toprule    
        \textbf{Dataset} and \textbf{Task} & \multicolumn{2}{c}{MAE ($\downarrow$)}  & \multicolumn{2}{c}{Training time (seconds)} & \multicolumn{2}{c}{Inference time (seconds)}\\
        \cmidrule(l{2pt}r{2pt}){2-3}
        \cmidrule(l{2pt}r{2pt}){4-5}
        \cmidrule(l{2pt}r{2pt}){6-7} 
        & \rdl small & \our &  \rdl small & \our &  \rdl small & \our  \\
        \midrule
        \texttt{rel-hm} \textbf{item-sales} & 0.058 & \textbf{0.037} & 1909 & \textbf{83} & 3.12 &\textbf{ 0.16}\\
        \bottomrule
        \end{tabular}
    }
    \end{center}
    \label{tab:rdl_less_param_reg}
\end{table}

\section{Additional Details on Time Splitting}\label{app:splittime}
Our method involves using the Snapshotted Relational Graph at a specific time point. This time point can correspond to a day, a week, a month, or any chosen interval. We selected the interval length based on the window length of the validation set in each dataset of RelBench. The interval lengths for each dataset are summarized in \Cref{tab:time_intervals}.

\begin{table}[ht]            
        \caption{Selected time intervals for each dataset.}
        \begin{center}
    \scriptsize
        \begin{tabular}{lr}
        \toprule
        \textbf{Dataset} & \textbf{t} \\
        \midrule
        \texttt{rel-f1}& 1 month    \\
        \texttt{rel-hm}               & 7 days     \\
        \texttt{rel-stack}           & 3 months   \\
        \texttt{rel-event}            & 7 days     \\
        \texttt{rel-amazon}           & 3 months   \\
        \bottomrule
        \end{tabular}
        \label{tab:time_intervals}
        \end{center}
\end{table}

\section{\our \textsc{w/o time} with Feature Engineering}\label{app:our_no_time_with_FE}
We wanted to test the hypothesis that even when generalist models are allowed to use these engineered features, the performance remains suboptimal, underscoring the fact that such features are specifically tailored for tabular models. We conducted a preliminary experiment on the \textbf{driver-top3} task of the \texttt{re-f1} dataset, where the same engineered features were directly applied as node features in the graph without employing any tabular methods (\our \xspace with F.E.) (\Cref{tab:feongraph}). The poor performance confirms that the feature produced by the feature engineering are primarily designed for tabular models.

\begin{table}[t]
\scriptsize
\caption{\our \textsc{w/o time} with F.E. performs worse than both \our \textsc{w. P.} and \our, indicating that GNNs do not benefit from the feature engineering, which was specifically designed for tabular models.}
\begin{center}
    \begin{tabular}{lc}
    \toprule
        & \texttt{rel-f1} \textbf{driver-top3} \\
         \cmidrule(l{2pt}r{2pt}){2-2}
        & Test \\    
    \midrule
    \our \textsc{w/o time} 
                        & \valstd{82.28}{2.44}\\
    \our\xspace \textsc{w/o time} with F.E.
                        & \valstd{82.58}{2.40}\\
    \our \textsc{w. P.}
                        & \valstd{83.28}{0.73}\\
    \our
                        & \valstdf{84.73}{1.43}\\
    \bottomrule
    \end{tabular}\label{tab:feongraph}
    \end{center}
\end{table}

\section{Experimental Setup and Reproducibility}\label{app:Exp_setup_and_rep}
In this section, we provide a detailed description of the model architectures used in this work, along with the implementation details required to ensure reproducibility. \our and its variants (\our \textsc{w/o time} and \our \textsc{w. P.}) are constructed by stacking heterogeneous GraphSAGE layers~\cite{fey2019fast,hamilton2017inductive}, employing sum-based neighbor aggregation.

The hyperparameter search was performed using grid search, exploring values for the learning rate ($0.1, 0.01, 0.001$), dropout rates ($0.1, 0.2, 0.3, 0.4, 0.5$), hidden dimensions ($16,32,64,128$), and the number of layers (ranging from $2$ to $6$). Our method is implemented with PyTorch, PyTorch Geometric~\citep{fey2019fast}, and TorchFrame~\citep{hu2024pytorch}, and the experiments were conducted on a single RTX-4090 GPU with 24GB of memory. For classification tasks, we used the BCEWithLogitsLoss function, and for regression tasks, we employed L1Loss. The source code for reproducibility is available at \url{https://github.com/AntonioLonga/LightRDL}.

\changed{We also performed a comprehensive hyperparameter search for \xgb, selecting the configuration that achieved the best performance on the validation set. The search space for each hyperparameter is detailed in \Cref{tab:xgb-hyperparams}. Logarithmic scales were applied to the learning rate as well as the L1 and L2 regularization terms. Specifically, \texttt{max\_depth} controls the depth of each tree, \texttt{num\_leaves} determines the maximum number of leaves per tree, \texttt{subsample} and \texttt{colsample\_bytree} define the sampling ratios for training instances and features respectively, and \texttt{min\_data\_in\_leaf} sets the minimum number of samples required in a leaf.}

\begin{table}[t]
\scriptsize
\centering
\caption{\changed{Hyperparameter search space for \xgb.}}
\label{tab:xgb-hyperparams}
\begin{tabular}{ll}
\toprule
\textbf{Hyperparameter}        & \textbf{Search Space} \\
\midrule
\texttt{max\_depth}            & [3, 11] \\
\texttt{learning\_rate}        & [$10^{-3}$, 0.1] (log scale) \\
\texttt{num\_leaves}           & [2, 1024] \\
\texttt{subsample}            & [0.05, 1.0] \\
\texttt{colsample\_bytree}     & [0.05, 1.0] \\
\texttt{lambda\_l1}            & [$10^{-9}$, $10^{1}$] (log scale) \\
\texttt{lambda\_l2}            & [$10^{-9}$, $10^{1}$] (log scale) \\
\texttt{min\_data\_in\_leaf}   & [1, 100] \\
\bottomrule
\end{tabular}
\end{table}

\section{Distillation details}\label{app:distillation_details}
The knowledge distillation from \xgb into an MLP was carried out as described in Section \ref{par:dist}. A grid search was performed to tune the learning rate, dropout, $\alpha$ value, temperature $T$, and number of layers. The size of the penultimate layer, from which the embeddings are extracted, was fixed at 10. The results of the distillation for the classification tasks are presented in \Cref{tab:resdistclas}, where the first column shows the ROCAUC against the true target, and the second column shows the ROCAUC against the predictions of \xgb. In \Cref{tab:resdistreg}, the results for the regression tasks are also reported.

\begin{table}[t]
\scriptsize
\caption{Distillation results for the regression tasks in MAE with respect to the real target and to the \xgb prediction.}
    \begin{center}
    \begin{tabular}{lrr}
    \toprule
    \textbf{Dataset} and \textbf{Task} & \textbf{MAE vs Real} & \textbf{MAE vs \xgb} \\ 
    \midrule
    \texttt{rel-f1}  \textbf{driver-position}   & 3.881              & 2.411\\
    \texttt{rel-hm}  \textbf{item-sales}        & 0.040              & 0.022\\
    \texttt{rel-event} \textbf{user-atten.} & 0.269              & 0.068\\
    \texttt{rel-stack}  \textbf{post-votes}     & 0.067              & 0.007\\
    \texttt{rel-amazon} \textbf{user-ltv}   & 14.438  & 7.319\\ 
    \texttt{rel-amazon} \textbf{item-ltv}     & 50.925 & 32.264 \\ 
    \bottomrule
    \end{tabular}\label{tab:resdistreg}
    \end{center}
\end{table}

\begin{table}[t]
\scriptsize
\caption{Distillation results for the classification tasks in ROCAUC with respect to the real target and to the \xgb prediction.}
    \begin{center}
    \begin{tabular}{lrr}
    \toprule
    \textbf{Dataset} and \textbf{Task} & \textbf{ROCAUC vs Real} & \textbf{ROCAUC vs \xgb} \\ 
    \midrule
        \texttt{rel-f1}  \textbf{driver-top3}   & 82.74  & 89.92\\ 
        \texttt{rel-f1}  \textbf{driver-dnf}     & 79.32& 90.17\\
        \texttt{rel-hm}  \textbf{user-churn}     & 69.81 & 82.03\\
        \texttt{rel-event} \textbf{user-ignore} & 80.23 & 91.08\\
        \texttt{rel-event}  \textbf{user-badge}     & 85.77 & 92.01 \\
        \texttt{rel-stack}  \textbf{user-engage} & 87.05 & 89.41\\ 
        \texttt{rel-amazon}  \textbf{user-churn}   & 66.93  & 89.32\\ 
        \texttt{rel-amazon}  \textbf{item-churn}     & 79.91 & 89.73 \\ 
    \bottomrule
    \end{tabular}\label{tab:resdistclas}
    \end{center}

\end{table}

\section{Window Size Sensitivity}
\label{app:varing_windows_size}
\changed{To evaluate the sensitivity of our method to the choice of time window size $t$, we conducted an additional experiment varying $t$ on the three node-level tasks (\texttt{driver-position}, \texttt{driver-dnf}, \texttt{driver-top3}) from the RelBench benchmark. The default window size used in our experiments is 30 days. We tested smaller and larger windows ranging from 13 days (the shortest reasonable window, considering races occur every 12 days) up to 60 days.}

\changed{\textbf{Performance Robustness.} As shown in \Cref{tab:window_perf}, model performance remains stable across all window sizes, with variations well within one standard deviation. This confirms the robustness of our approach with respect to $t$, indicating that it is not overly sensitive to the specific choice of time window.}

\changed{\textbf{Training Time.} As expected, increasing the window size leads to larger snapshot graphs, which slightly increases training time per epoch. However, the increase is moderate and does not significantly impact model efficiency, as detailed in \Cref{tab:window_time}.}

\begin{table}[t]
\scriptsize
\caption{\changed{Performance on the \texttt{rel-f1} node-level tasks across different time window sizes $t$. Performance is reported as test MAE or ROC-AUC ± standard deviation.}}
\begin{center}
\begin{tabular}{lccccc}
\toprule
\textbf{Task} & \textbf{13 days} & \textbf{20 days} & \textbf{30 days} & \textbf{45 days} & \textbf{60 days} \\
\midrule
\texttt{driver-position} (MAE)    & 3.85 (0.03) & 3.86 (0.03) & 3.86 (0.04) & 3.88 (0.02) & 3.99 (0.02) \\
\texttt{driver-dnf} (ROCAUC) & 73.0 (0.2)  & 73.0 (0.3)  & 73.5 (0.3)  & 73.6 (0.4)  & 73.3 (0.3) \\
\texttt{driver-top3} (ROCAUC) & 84.5 (0.4)  & 84.1 (0.4)  & 84.7 (0.4)  & 85.1 (0.4)  & 84.9 (0.3) \\
\bottomrule
\end{tabular}
\label{tab:window_perf}
\end{center}
\end{table}

\begin{table}[t]
\scriptsize
\caption{\changed{Training time per epoch (in seconds) across different time window sizes $t$ on the \texttt{rel-f1} tasks.}}
\begin{center}
\begin{tabular}{lccccc}
\toprule
\textbf{Task} & \textbf{13 days} & \textbf{20 days} & \textbf{30 days} & \textbf{45 days} & \textbf{60 days} \\
\midrule
\texttt{driver-position}   & 1.40 & 1.50 & 1.50 & 1.60 & 1.70 \\
\texttt{driver-dnf}    & 0.12 & 0.16 & 0.16 & 0.18 & 0.21 \\
\texttt{driver-top3}   & 0.08 & 0.09 & 0.09 & 0.12 & 0.16 \\
\bottomrule
\end{tabular}
\label{tab:window_time}
\end{center}
\end{table}

\section{Training Time Per Epoch}\label{app:train_time_epoch}

In this section, we analyze and compare the training time per epoch for \our and \rdl. Table \ref{tab:training_time_epoch} reports the detailed measurements. On average, \our is 423 times faster than \rdl, with the maximum observed speedup reaching an impressive 2898 times.

\begin{table}[t]
    \scriptsize
        \caption{Training time per epoch shows \our is on average 423$\times$ faster than rdl in training.}
        \begin{center}
        \begin{tabular}{llrr>{\columncolor{mycustomcolor2}}r}
            \toprule
            & \textbf{Dataset} and \textbf{Task} &  \rdl   & \textbf{\our} & Training Speedup $(\uparrow)$ w.r.t.\\
            \midrule
            \multirow{6}{*}{\centering \rotatebox[origin=c]{90}{Regression} }
            &   \texttt{rel-f1} \textbf{driver-position} &   58.38      & 1.53 & \textbf{116}\\
            &   \texttt{rel-hm} \textbf{item-sales} &   2040.56         & 0.10 & \textbf{1333}\\
            &   \texttt{rel-event} \textbf{user-atten.}  & 9.84     & 7.00 & \textbf{101}\\
            &   \texttt{rel-stack} \textbf{post-votes} & 1545.65        & 0.96 & \textbf{221}\\
            &   \texttt{rel-amazon} \textbf{user-ltv} &   16.22         & 1.05 & \textbf{17}\\
            &   \texttt{rel-amazon} \textbf{item-ltv} &  18.54          & 1.05 & \textbf{18}\\
            \midrule
            \multirow{9}{*}{\centering \rotatebox[origin=c]{90}{Classification}}
            &   \texttt{rel-f1} \textbf{driver-dnf}  &    30.34         & 0.16 & \textbf{190}\\
            &   \texttt{rel-f1} \textbf{driver-top3} &    11.27         & 0.09 & \textbf{119}\\
            &   \texttt{rel-hm} \textbf{user-churn}  &  378.32          & 0.13 & \textbf{2898}\\
            &   \texttt{rel-event} \textbf{user-ignore} & 9.59          & 0.08 & \textbf{120}\\
            &   \texttt{rel-event} \textbf{user-repeat} &   2.24        & 0.02 & \textbf{100}\\
            &   \texttt{rel-stack} \textbf{user-engage} &   942.58  & 3.22 & \textbf{293}\\
            &   \texttt{rel-stack} \textbf{user-badge} &  3579.12       & 4.22 & \textbf{849}\\ 
            &   \texttt{rel-amazon} \textbf{user-churn} &   15.22       & 0.71 & \textbf{21}\\
            &   \texttt{rel-amazon} \textbf{item-churn} &   11.32       & 0.70 & \textbf{16}\\
            \midrule
            \multicolumn{2}{r}{\textbf{avg.}} & 577.95 & 1.36 & \textbf{423}$\times$ faster\\
            \bottomrule
        \end{tabular}
        \end{center}
        \label{tab:training_time_epoch}
    \end{table}

\subsection{Impact of Batch Size on Training Time}

When comparing the training time per epoch, it is essential to consider the role of batch size, as it significantly affects both computational efficiency and memory usage. 

For \our, a batch size of 1 encompasses the entire graph at a given timestamp, allowing the model to process all interactions occurring at that time simultaneously. In contrast, for \rdl, a batch size of 1 refers to a single node and its temporal neighborhood, which represents a much smaller computational unit.

Given these fundamental differences in how batch size is defined for the two models, a direct comparison would be unfair without proper adjustments. To ensure a balanced evaluation, we manually tuned the batch sizes for both models to achieve comparable GPU memory usage. This adjustment ensures that the observed differences in training time are not biased by resource allocation disparities. The results reported in \Cref{tab:comparison} and \Cref{fig:time} were obtained by strictly following the procedure just described.

Table \ref{tab:batchsize} summarizes the batch sizes and memory consumption for both models, providing a comprehensive overview of the setup used for the training time analysis.

\begin{table}[t]
\scriptsize
    \caption{Batch size comparison: both \rdl and our \our are trained using batch sizes that utilize the same amount of memory.}
    \begin{center}
    \begin{tabular}{llrrrr}
        \toprule
        & \textbf{Dataset} and \textbf{Task} & \rdl & \textbf{\our} & \rdl & \textbf{\our}  \\
        \cmidrule(l{2pt}r{2pt}){3-4} \cmidrule(l{2pt}r{2pt}){5-6}
        & &   \multicolumn{2}{c}{batch size} & \multicolumn{2}{c}{Memory usage \changed{(MB)}}   \\
        \midrule
        \multirow{6}{*}{\centering \rotatebox[origin=c]{90}{Regression} }
            &  \texttt{rel-f1}  \textbf{driver-position} &  64 & All & 1548 & 1484 \\
            &  \texttt{rel-hm}  \textbf{item-sales} & 32 & All & 6874 & 6502 \\
            &  \texttt{rel-event}  \textbf{user-atten.} & 128 & All & 5524 & 5466 \\
            &  \texttt{rel-stack}  \textbf{post-votes} & 256 & 1 & 15214 & 15122 \\
            &  \texttt{rel-amazon}  \textbf{user-ltv} & 1024 & 1 & 19125 & 19326 \\
            &  \texttt{rel-amazon}  \textbf{item-ltv} & 1024 & 1 & 19584 & 19548 \\
        \midrule
        \multirow{9}{*}{\centering \rotatebox[origin=c]{90}{Classification} }
            & \texttt{rel-f1}  \textbf{driver-dnf} & 64 & All & 1538 & 1441 \\
            & \texttt{rel-f1}  \textbf{driver-top3} & 64 & All & 1508 & 1450 \\
            & \texttt{rel-hm}  \textbf{user-churn}  & 32 & All & 6892 & 6512 \\
            & \texttt{rel-event}  \textbf{user-ignore} & 128 & All & 5324 & 5364 \\
            & \texttt{rel-event}  \textbf{user-repeat} & 128 & All & 5622 & 5536 \\
            & \texttt{rel-stack}  \textbf{user-eng.} & 256 & 1 & 15234 & 14508 \\
            & \texttt{rel-stack}  \textbf{user-badge} & 256 & 1 & 15058 & 14556 \\
            & \texttt{rel-amazon}  \textbf{user-churn} & 1024 & 1 & 22125 & 21554 \\
            & \texttt{rel-amazon}  \textbf{item-churn} & 1024 & 1 & 22844 & 21589 \\
        \bottomrule
    \end{tabular}
    \end{center}
    \label{tab:batchsize}
\end{table}

\section{Additional Results on Inference Time}\label{app:inftime}
In \cref{app:table:inference_time}, we present the inference times for \our model, \rdl, and \xgb. Additionally, we highlight the relative speedup gains with a yellow background. The results demonstrate a significant speedup compared to \rdl and a slight improvement in speed over \xgb.

\begin{table*}[t]
    \caption{Inference time (seconds) for \our compared to \rdl and \xgb. \our is significantly faster than \rdl, and is slightly faster than \xgb.}
    \begin{center}
    \scriptsize
    \begin{tabular}{p{0.3mm}lcrr>{\columncolor{mycustomcolor2}}r>{\columncolor{mycustomcolor2}}r}
        \toprule
         &   & && \textbf{\our}& \multicolumn{2}{c}{\cellcolor{mycustomcolor2} Inference Speedup $(\uparrow)$ w.r.t.}\\
         & \textbf{Dataset} and \textbf{Task} &\xgb & \rdl &  Distillation + R-GNN & \xgb & \rdl \\ 
        \midrule
        \multirow{6}{*}{\rotatebox[origin=c]{90}{Regression}}&
        \texttt{rel-f1} \textbf{driver-position} 
            & 0.04 & 2.34 & 0.01 + 0.03 = 0.04 & 1.0 &\textbf{29.3}\\
             & \texttt{rel-hm} \textbf{item-sales} 
            & 0.13 & 10.52 & 0.01 + 0.01 = 0.02 & \textbf{6.5} & \textbf{65.8}  \\
            & \texttt{rel-event} \textbf{user-atten.} 
            & 0.04 & 0.95 & 0.01 + 0.01 = 0.02 & \textbf{2.0} &\textbf{11.9} \\
            & \texttt{rel-stack} \textbf{post-votes} 
            & 0.79 & 35.23 & 0.05 + 0.23 = 0.28 & \textbf{2.8} & \textbf{33.6}\\
            & \texttt{rel-amazon} \textbf{user-ltv} 
            & 0.14 & 5.30 & 0.01 + 0.08 = 0.09 &\textbf{1.6} & \textbf{22.1}  \\
            & \texttt{rel-amazon} \textbf{item-ltv} 
            & 0.04 & 5.48 & 0.01 + 0.03 = 0.04 & 1.0 & \textbf{45.7} \\
        \midrule   

        \multirow{9}{*}{\rotatebox[origin=c]{90}{Classification}}
            & \texttt{rel-f1} \textbf{driver-dnf} 
            & 0.05 & 1.79 & 0.01 + 0.02 = 0.03 & \textbf{1.7} & \textbf{22.4} \\
            &\texttt{rel-f1} \textbf{driver-top3} 
            & 0.04 & 2.19 & 0.01 + 0.03 = 0.04 & 1.0 & \textbf{27.4} \\
            &\texttt{rel-hm} \textbf{user-churn} 
            & 0.30 & 3.63 & 0.01 + 0.01 = 0.02 & \textbf{15.0} &\textbf{11.0} \\
            &\texttt{rel-event} \textbf{user-ignore} 
            & 0.02 & 1.19 & 0.01 + 0.01 = 0.02 & 1.0 & \textbf{2.4}  \\
            &\texttt{rel-event} \textbf{user-repeat} 
            & 0.04 & 2.29 & 0.01 + 0.01 = 0.02 & \textbf{2.0} &\textbf{32.7} \\
            &\texttt{rel-stack} \textbf{user-eng.} 
            & 0.14 & 24.01 & 0.05 + 0.25 = 0.30 & 0.5&\textbf{57.2} \\
            &\texttt{rel-stack} \textbf{user-badge} 
            & 3.03 & 96.23 & 0.05 + 0.27 = 0.32 & \textbf{9.5} &\textbf{28.3}  \\
            &\texttt{rel-amazon} \textbf{user-churn} 
            & 0.08 & 2.25 & 0.02 + 0.06 = 0.08 & 1.0&\textbf{12.5}  \\
            &\texttt{rel-amazon} \textbf{item-churn} 
            & 0.08 & 2.24 & 0.02 + 0.05 = 0.07 & \textbf{1.1} &\textbf{13.2}  \\
        \midrule
        \multicolumn{5}{r}{\textbf{avg.}}  & \textbf{3.2}$\times$ faster & \textbf{124.3}$\times$ faster \\
        \bottomrule
    \end{tabular}
    \end{center}
    \label{app:table:inference_time}
\end{table*}

\section{Additional Results on Performance}\label{app:res_with_std}
In \cref{app:tab:teststd}, we show the test performance along with the standard deviation, averaged across five runs. Similarly, \cref{app:tab:validationstd} presents the validation performance with the corresponding standard deviation, also averaged over five runs.

\begin{table*}[t]
    \caption{Comparison of performance metrics for \xgb,  \rdl and our \our across multiple tasks on test set.}\label{app:tab:teststd}
    \begin{center}
    \scriptsize
    \begin{tabular}{lrrr}
        \toprule
        & \multicolumn{3}{c}{Regression (MAE $\downarrow$)}\\
         \cmidrule{2-4}
        \textbf{Dataset} and \textbf{Task} & \xgb & \rdl & \textbf{\our} \\
        \midrule
         \texttt{rel-f1} \textbf{driver-position} & 4.01{\tiny ($\pm 0.08$)} & 4.142{\tiny ($\pm 0.11$)} & 3.861{\tiny ($\pm 0.045$)} \\
         \texttt{rel-hm} \textbf{item-sales} & 0.038{\tiny ($\pm 0.001$)} & 0.056{\tiny ($\pm 0.001$)} & 0.037{\tiny ($\pm 0.001$)} \\
         \texttt{rel-event} \textbf{user-atten.} & 0.249{\tiny ($\pm 0.003$)} & 0.255{\tiny ($\pm 0.004$)} & 0.238{\tiny ($\pm 0.003$)} \\
         \texttt{rel-stack} \textbf{post-votes} & 0.068{\tiny ($\pm 0.011$)} & 0.065{\tiny ($\pm 0.021$)} & 0.064{\tiny ($\pm 0.008$)} \\
         \texttt{rel-amazon} \textbf{user-ltv} & 14.212{\tiny ($\pm 0.002$)} & 14.314{\tiny ($\pm 0.013$)} & 13.587{\tiny ($\pm 0.004$)} \\
         \texttt{rel-amazon} \textbf{item-ltv} & 49.917{\tiny ($\pm 0.003$)} & 50.053{\tiny ($\pm 0.163$)} & 48.112{\tiny ($\pm 0.006$)} \\
        \bottomrule
        \\
        \toprule
        & \multicolumn{3}{c}{ Classification  (ROCAUC $\uparrow$)}\\
        \cmidrule{2-4}
        \textbf{Dataset} and \textbf{Task} & \xgb & \rdl & \textbf{\our} \\
        \midrule
         \texttt{rel-f1} \textbf{driver-dnf} & 70.52{\tiny ($\pm 1.07$)} & 71.08{\tiny ($\pm 2.79$)} & 73.55{\tiny ($\pm 0.34$)} \\
         \texttt{rel-f1} \textbf{driver-top3} & 82.77{\tiny ($\pm 1.08$)} & 80.30{\tiny ($\pm 1.85$)} & 84.73{\tiny ($\pm 0.43$)} \\
         \texttt{rel-hm} \textbf{user-churn} & 69.12{\tiny ($\pm 0.01$)} & 69.09{\tiny ($\pm 0.35$)} & 68.93{\tiny ($\pm 0.03$)} \\
         \texttt{rel-event} \textbf{user-ignore} & 82.62{\tiny ($\pm 1.14$)} & 77.82{\tiny ($\pm 1.88$)} & 83.98{\tiny ($\pm 0.44$)} \\
         \texttt{rel-event} \textbf{user-repeat} & 75.78{\tiny ($\pm 1.74$)} & 76.50{\tiny ($\pm 0.78$)} & 77.77{\tiny ($\pm 0.68$)} \\
         \texttt{rel-stack} \textbf{user-eng.} & 90.34{\tiny ($\pm 0.09$)} & 90.59{\tiny ($\pm 0.03$)} & 89.02{\tiny ($\pm 0.03$)} \\
         \texttt{rel-stack} \textbf{user-badge} & 86.34{\tiny ($\pm 0.04$)} & 88.54{\tiny ($\pm 0.15$)} & 86.71{\tiny ($\pm 0.53$)} \\
         \texttt{rel-amazon} \textbf{user-churn} & 68.34{\tiny ($\pm 0.09$)} & 70.42{\tiny ($\pm 0.05$)} & 69.87{\tiny ($\pm 0.19$)} \\
         \texttt{rel-amazon} \textbf{item-churn} & 82.62{\tiny ($\pm 0.03$)} & 82.81{\tiny ($\pm 0.03$)} & 83.84{\tiny ($\pm 0.08$)} \\
        \bottomrule
    \end{tabular}
    \end{center}
\end{table*}

\begin{table*}[t]
    \caption{Comparison of performance metrics for \xgb,  \rdl and our \our across multiple tasks on validation set.}\label{app:tab:validationstd}
    \begin{center}
    \scriptsize
    \begin{tabular}{lrrr}
        \toprule
        & \multicolumn{3}{c}{Regression (Validation MAE $\downarrow$)}\\
        \cmidrule{2-4}
        \textbf{Dataset} and \textbf{Task} & \xgb & \rdl & \textbf{\our} \\
        \midrule
         \texttt{rel-f1} \textbf{driver-position} & 2.80{\tiny ($\pm 0.030$)} & 3.13{\tiny ($\pm 0.050$)} & 2.910{\tiny ($\pm 0.070$)} \\
         \texttt{rel-hm} \textbf{item-sales} & 0.049{\tiny ($\pm 0.001$)} & 0.065{\tiny ($\pm 0.001$)} & 0.0458{\tiny ($\pm 0.000$)} \\
         \texttt{rel-event} \textbf{user-atten.} & 0.249{\tiny ($\pm 0.002$)} & 0.246{\tiny ($\pm 0.004$)} & 0.2441{\tiny ($\pm 0.002$)} \\
         \texttt{rel-stack} \textbf{post-votes} & 0.062{\tiny ($\pm 0.001$)} & 0.059{\tiny ($\pm 0.001$)} & 0.0587{\tiny ($\pm 0.001$)} \\
         \texttt{rel-amazon} \textbf{user-ltv} & 11.483{\tiny ($\pm 0.001$)} & 12.13{\tiny ($\pm 0.007$)} & 11.325{\tiny ($\pm 0.025$)} \\
         \texttt{rel-amazon} \textbf{item-ltv} & 44.315{\tiny ($\pm 0.002$)} & 45.14{\tiny ($\pm 0.068$)} & 43.121{\tiny ($\pm 0.078$)} \\
        \bottomrule
        \\
        \toprule
        & \multicolumn{3}{c}{ Classification  (Validation ROCAUC $\uparrow$)}\\
        \cmidrule{2-4}
        \textbf{Dataset} and \textbf{Task} & \xgb & \rdl & \textbf{\our} \\
        \midrule
         \texttt{rel-f1} \textbf{driver-dnf} & 81.49{\tiny ($\pm 0.25$)} & 75.19{\tiny ($\pm 2.64$)} & 81.90{\tiny ($\pm 0.77$)} \\
         \texttt{rel-f1} \textbf{driver-top3} & 89.74{\tiny ($\pm 0.25$)} & 76.25{\tiny ($\pm 2.22$)} & 89.15{\tiny ($\pm 0.33$)} \\
         \texttt{rel-hm} \textbf{user-churn} & 70.01{\tiny ($\pm 0.02$)} & 69.82{\tiny ($\pm 0.33$)} & 69.30{\tiny ($\pm 0.04$)} \\
         \texttt{rel-event} \textbf{user-ignore} & 91.08{\tiny ($\pm 1.61$)} & 90.66{\tiny ($\pm 0.50$)} & 91.84{\tiny ($\pm 0.11$)} \\
         \texttt{rel-event} \textbf{user-repeat} & 73.18{\tiny ($\pm 0.44$)} & 72.56{\tiny ($\pm 0.79$)} & 74.52{\tiny ($\pm 0.46$)} \\
         \texttt{rel-stack} \textbf{user-eng.} & 87.84{\tiny ($\pm 0.02$)} & 89.62{\tiny ($\pm 0.13$)} & 88.56{\tiny ($\pm 0.21$)} \\
         \texttt{rel-stack} \textbf{user-badge} & 90.17{\tiny ($\pm 0.03$)} & 90.19{\tiny ($\pm 0.05$)} & 89.88{\tiny ($\pm 0.02$)} \\
         \texttt{rel-amazon} \textbf{user-churn} & 68.79{\tiny ($\pm 0.02$)} & 70.45{\tiny ($\pm 0.06$)} & 72.01{\tiny ($\pm 0.12$)} \\
         \texttt{rel-amazon} \textbf{item-churn} & 82.41{\tiny ($\pm 0.02$)} & 82.39{\tiny ($\pm 0.02$)} & 83.11{\tiny ($\pm 0.04$)} \\
        \bottomrule
    \end{tabular}
    \end{center}
\end{table*}

\end{document}